# Deformation of upper mantle rocks with contrasting initial fabrics in axial extension

Walid Ben Ismail[1,#], Andréa Tommasi[2,*], Marco A. Lopez-Sanchez[2], Ernest H. Rutter[1], Fabrice Barou[2], Sylvie Demouchy[2]

[1]*Rock Deformation Laboratory, Dept. Earth & Environmental Sciences, University of Manchester, UK*

[2]*Géosciences Montpellier, CNRS & Université de Montpellier, Montpellier, France*

*\* Corresponding author: andrea.tommasi@umontpellier.fr*

[#] *Now at Schlumberger Oilfield Services, The Woodland, Texas, USA*

*Abstract*

To explore the role of the olivine grain size and crystal preferred orientation (CPO), on the evolution of the microstructure and the mechanical behaviour of upper mantle rocks up to large strains, we performed axial extension experiments at 1200°C, 300MPa confining pressure, and constant displacement rate on three natural peridotites: a fine-grained mylonitic harzburgite with a weak olivine CPO and two coarse-grained well-equilibrated dunites with olivine CPO of variable intensity. Despite the contrasting microstructures, initial flow stresses show a limited range of variation (115-165±5MPa), with the fine-grained harzburgite displaying the highest initial strength. However, the evolution of both mechanical behavior and microstructure differs markedly between fine and coarse-grained peridotites. In the fine-grained harzburgite, necking is associated with decrease in the apparent differential stress. Focusing of strain and stress resulted in increase of the olivine CPO intensity and recrystallized fraction and decrease of the recrystallized grain size in the neck. Analysis of the final stress and strain in the neck indicates softening in response to the evolution of the microstructure and CPO. In contrast, necking of the coarse-grained dunite samples is associated with a weaker or no decrease in the apparent differential stress. This implies hardening, consistently with the increase in bulk intragranular misorientation with increasing strain observed in these samples and final stresses in the neck similar or higher than the initial ones. Coarse-grained dunites deformed heterogeneously. Crystals well oriented to deform by dislocation glide became elongated and developed undulose extinction, whereas crystals in hard orientations remained almost undeformed. In the neck, stress and strain concentration resulted in formation of kinks in "hard" crystals and dynamic recrystallization in "soft" crystals. We interpret the more effective strain-induced softening of the fine-grained peridotite as due to easier dynamic recrystallization, probably due to the higher proportion of grain boundaries acting as nucleation sites.

# 1. Introduction

The deformation processes and, by consequence, the rheology of mantle rocks depend on their microstructure, particularly: (1) the grain size, which controls the relative contribution of grain interiors and boundaries to the deformation (Hirth and Kohlstedt, 2003; Nicolas and Poirier, 1976) and (2) the orientation of the olivine crystals relative to the imposed stresses, since crystal preferred orientation (CPO) may result in significant viscoplastic anisotropy (Hansen et al., 2016, 2012; Mameri et al., 2019; Tommasi et al., 2009; Tommasi and Vauchez, 2001). In nature, both parameters are extremely variable and evolve in response to local deformation and thermal histories. Grain sizes in naturally deformed peridotites are dominantly millimetric, but they may range from a few microns in ultramylonites (e.g., Hidas et al., 2016; Kaczmarek and Tommasi, 2011; Newman et al., 1999) to centimeters in cratonic xenoliths (e.g., Baptiste et al., 2012; Vauchez et al., 2005) and in dunites deformed in presence of melts (e.g., Higgie and Tommasi, 2012; Tommasi et al., 2017). Olivine CPO intensities also vary from very weak (J-indexes ≤2) to very strong (J-indexes >10) with a peak at moderate strength CPO (J-indexes=4-5; Tommasi and Vauchez, 2015).

Most rheological data for olivine-rich rocks have been derived from axial compression experiments (e.g., Chopra and Paterson, 1981; Faul et al., 2011; Karato et al., 1986; Thieme et al., 2018; Zhao et al., 2009). At low strains, these experiments allow a precise determination of stress and strain as a function of time, resulting in reliable flow laws. However, beyond 25% shortening, barreling occurs and the deformation becomes inhomogeneous. This prevents the investigation of the effect of an evolving microstructure on the rheological behavior. Moreover, to ensure reproducibility, most experiments were performed on synthetic olivine polycrystals with random CPO and fine-grain sizes (5-30 µm). Rare axial compression experiments up to high strains (≤71%) on medium-grained natural dunites (Aheim dunite with initial grain sizes ~350 µm and moderate olivine CPO) showed different evolutions of the CPO in samples compressed parallel, normal, or oblique to the preexisting foliation (Boneh and Skemer, 2014). However, a detailed analysis of the evolution of the microstructure and its influence on the mechanical behavior is unavailable.

Development of deformation experiments in simple shear (Zhang et al., 2000; Zhang and Karato, 1995) and torsion (Bystricky et al., 2000; Demouchy et al., 2012; Hansen et al., 2014, 2012) allowed deformation of fine-grained synthetic olivine aggregates up to high shear strains (up to 20 in torsion tests). These experiments constrained the evolution of the microstructure and olivine CPO during simple shear and of the effect of this evolution on the mechanical behavior. They showed that dynamic recrystallization first accelerates the rotation of olivine crystals towards easy glide orientations and then stabilizes the CPO (Bystricky et al., 2000; Hansen et al., 2014), an observation that was corroborated by numerical modelling (Signorelli and Tommasi, 2015). Torsion tests also systematically display strain softening, probably due to both evolution of the CPO and dynamic recrystallization. However, exploring the influence of the initial microstructure via torsion experiments is a challenge, since a high strength of the samples produces decoupling of the sample ends from the rotating piston. A few successful torsion experiments on coarse-grained dunites up to shear strains of 3.9 were, nevertheless, achieved by using 'dog-bone' shaped samples (Skemer et al., 2011). In these non-conventional torsion experiments, the wide ends of the sample provided sufficient contact area to ensure coupling and support moderate torques, which produced large stresses in the narrower center of the sample due to the strong radial dependence between shear stress and torque in a cylindrical geometry.

Axisymmetric extension experiments also allow attainment of large strains. They are widely used for testing of ductile metals and, although less frequent, have also been successfully applied to rocks (e.g., Brodie and Rutter, 2000; Hart, 1967; Heard, 1963; Rutter, 1998, 1995). Extension tests on synthetic aggregates of fine-grained iron-rich olivine (Fo50) were reported by Hansen et al. (2016), who associated extension and torsion tests to quantify the viscoplastic anisotropy due to an olivine CPO. In axial extension, the bulk strain may exceed 50%. However, the strain distribution in the sample is often heterogeneous, as geometric instability produces localized necking of the specimen if strain-rate softening due to the non-linearity of the flow law is not counteracted by microstructural hardening (Hart, 1967; Rutter, 1998). Both stress and strain rate vary along the length of a necked sample. For flow at constant volume, the local axial strain is equal to the strain corresponding to the local reduction of area normal to the stretching direction. The stress distribution may also be reconstructed from the variation of the sectional area of the sample (Rutter, 1998). Microstructures and CPO formed at various stresses, strain rates, and finite strains are therefore preserved within a single sample, making extension tests particularly well suited to study the evolution of the microstructure and CPO as a function of strain and stress, in the same way as torsion tests performed on full cylinders.

In this article, we examine the mechanical behavior and the evolution of the microstructure and CPO of olivine as a function of the finite strain in a series of extension experiments performed on three natural peridotites with varied initial microstructures and olivine CPO. By comparing their mechanical behavior and microstructural evolution, we analyze the effect of the grain size and CPO on the peridotites strength and active deformation mechanisms and characterize the interplay between viscoplastic deformation and dynamic recrystallization on the evolution of the mechanical behavior.

## 2. Methods

Thirteen axial extension experiments (Table 1) were performed at the Rock Deformation Laboratory, University of Manchester, UK, using a high resolution gas-medium high-pressure, internally-heated triaxial deformation apparatus (Paterson, 1990) modified for extension testing by incorporating a bayonet connector between the bottom loading piston and the top of the internal axial load cell.

Cylinders of 10 mm diameter and 21-27 mm long were cored in different orientations relative to the preexisting fabric in three starting materials (Table 1). After coring, the cylinders were stored in an oven at 80 °C until required for use. For the experiments, the samples were enclosed in a double iron jacket, the inner one capped by iron discs, which aimed at setting the oxygen fugacity at the Fe-FeO buffer. Silica activity was buffered by the natural presence of orthopyroxene in all samples.

All tests were performed at 1200 °C and 300 MPa confining pressure of argon. Temperature is accurate by ±5 °C over a profile longer than the sample, avoiding temperature heterogeneity as the sample is stretched. During confined extension tests, as those presented here, the sample ends do not need to be gripped because the confining pressure pushes the sample against the unloading piston as long as the axial differential load remains lower than the axial force due to the confining pressure. At the start of a test, the axial load is equal to that due to the confining pressure. During the experiment, withdrawing the piston reduces the axial load. All tests were unloaded at a constant rate of 2.5×10$^{-4}$ mm/s, corresponding to initial strain rates between 9.3×10$^{-6}$ and 1.2×10$^{-5}$ s$^{-1}$, depending on the length of the sample (Table 1). During the experiments, the strain rate slowly decreases due to lengthening of the samples (8.6×10$^{-6}$ to 1.1×10$^{-5}$ s$^{-1}$ at 10 % of bulk strain) until necking, which results in a heterogeneous strain distribution and high strain rates in the neck. The strain rate enhancement depends on the length of the most intensely deforming region, but strain rates at the neck are often more than five-fold the bulk ones in highly strained samples.

The resolution of the internal load cell is 50 N, resulting in an estimated accuracy of differential stress of ±5 MPa before necking (bulk strain < 20 %) and of ±15 MPa in the neck. Stress data was corrected for the contribution of the jacketing material (about 3% of the local stress value). Bulk differential stresses were calculated assuming homogeneous area reduction at constant volume along the length of the sample. Bulk strain was estimated as the overall change in length (after correction for axial apparatus stiffness) divided by original sample length. No attempt was made to correct the evolution of the bulk differential stresses for the effects of heterogenous deformation in the necked samples, because the *in situ* evolution of the shape of the sample over time cannot be assessed.

The distribution of differential stress and strain along the length of the specimen at the end of the experiment was calculated post-mortem by analyzing the variation of the area of the specimen normal to the extension direction (Hart, 1967; Rutter, 1998). We measured the diameter of each sample at 1 mm intervals along its length. As the axial force is equal at every point along the sample, the final axial force divided by the local area gives the local differential stress at the end of the experiment. Assuming deformation at constant volume, the local axial strain is deduced from the ratio between the actual area of the specimen normal to the extension direction and the area calculated assuming homogeneous deformation to the final bulk strain of the sample.

After the experiments, thin sections and polished blocks were cut parallel to the extension direction for optical observation of the microstructure and CPO measurement. Olivine and pyroxene crystallographic orientations were determined by indexation of electron back-scattered diffraction (EBSD) patterns generated on a field-emission scanning electron microscope – the Camscan Crystal Probe X500 FE at Geosciences Montpellier, equipped with the Oxford NordlysNano EBSD detector - and indexed using the AZtecHKL system. EBSD orientation maps of the entire recovered coarse peridotite samples, as well as of the starting material, were obtained with resolutions of 20 to 50 $\mu$m. In the highly strained samples, higher resolution (2 $\mu$m step) maps of the necking region were also made. For the fine-grained peridotite, two large-area maps were acquired: one in a low strain domain adjacent to the specimen end with a resolution of 2.5 $\mu$m and one at the center of the necking zone with a resolution of 1.5 $\mu$m.

EBSD data analysis, that is, calculation of the orientation distribution functions (ODF) and of the intragranular misorientations, plotting of pole figures, quantification of microstructural parameters (grain sizes, shape, and orientation) and of the geometrically necessary dislocation (GND) densities based on the intragranular misorientations using the Kernel Average Misorientation (KAM) and the Grain Orientation Spread (GOS) proxies, was performed using the MTEX toolbox v.5.2 in Matlab (http://mtex-toolbox.github.io/; Bachmann et al., 2011, 2010; Hielscher and Schaeben, 2008). ODFs were calculated using a "de la Vallée Poussin" kernel function with a half-width of 10°. Grain boundaries were defined with a misorientation threshold at 15°. Recrystallized grains were segmented based on the grain orientation spread (GOS < 2°). Local differential stresses were estimated based on the olivine piezometer of Van der Wal et al. (1993), which was calibrated on experiments using similar starting materials and deformation conditions as those in the present study. We used the GrainSizeTools script v3 (Lopez-Sanchez, 2018) to convert the mean apparent grain sizes expressed as equivalent circular diameters obtained using MTEX to linear intercept values, as in the original piezometer, and a stereological correction factor of 1.5.

Hydrogen concentration in olivine and orthopyroxene was measured in hand-polished slabs of the three starting materials, one annealed sample (VS14An), and two samples deformed to low bulk strains (VS14-3, VS15b3) by the analysis of hydroxyl groups in olivine using unpolarized transmission Fourier Transform Infrared (FTIR) spectroscopy. The details of the method are presented in the Supplementary Material. Hydrogen concentrations were calculated by integrating each spectrum over the 3620 – 3050 $cm^{-1}$ wavenumber range. Hydrogen concentrations for olivine were quantified using the calibrations of Paterson (1982) and Withers et al. (2012). These calibrations have a detection limit of about 0.5 ppm of $H_2O$ by weight for a 0.5 mm-thick olivine sample and an error around 30 % (Paterson 1982). For orthopyroxene, the calibrations of Paterson (1982) and Bell et al. (1995) were used.

## 3. Starting material

The starting material consisted of three naturally-deformed peridotites: a harzburgite (sample 98OA87) collected as a loose block derived from a km-scale low-temperature mylonitic zone in the Wuqba massif in the Oman ophiolite and two dunites (samples VS14 and VS15) sampled alongside the eastern boundary of the Balmuccia ultramafic massif in the Alps, at 45°50'15"N, 8°9'40"E.

The sample 98OA87 is a spinel-harzburgite (75% olivine Mg#=91; 20% opx; 3% cpx; 2% sp; area fractions based on EBSD mapping). It is a fine-grained mylonite with a well-developed lineation and foliation marked by the shape preferred orientation of olivine and (rare) enstatite porphyroclasts up to 1 mm long and by the orientation of finer-grained bands (Fig. 1a). Olivine porphyroclasts have aspect ratios varying from 1:2 up to 1:5. They show undulose extinction, closely spaced (100) subgrain walls, and irregularly shaped grain boundaries. The matrix is mainly composed of olivine grains 10 to 300 μm in diameter (median at 14 $\mu$m). It represents ca. 70% volume of the sample; the apparent area-weighted mean grain size in the matrix is 170 $\mu$m. Olivine displays a weak CPO characterized by poor alignment of [100] parallel to the lineation marked by the elongation of the porphyroclasts (Fig. 1a). The composition and microstructure of this sample are coherent with its origin in a depleted oceanic lithosphere that has been deformed under low-temperature conditions (<1000°C) in a shear zone normal to the ridge axis (Nicolas and Boudier, 1995). The sample lacks optically observable alteration, but FTIR analyses indicated minor, but widespread serpentinization along grain boundaries and fractures.

The two dunites display an annealed microstructure characterized by 1-3 mm wide polygonal grains of iron-rich olivine (Mg#=82-83) displaying low densities of intracrystalline deformation features, such as undulose extinction and subgrains (Fig. 1b,c). Enstatite and diopside (3-4% in VS14 and <2% in VS15; area fractions based on EBSD mapping) occur as isolated, irregularly-shaped interstitial grains, usually ≤1 mm in size, with no clear internal deformation features. The two dunites lack optically visible alteration (Fig. 1b,c) except for thin amphibole rims around the rare diopside crystals. However, FTIR analyses detected serpentinization along grain boundaries and fractures. Spinel (~1%) occurs as small rounded inclusions within olivine grains or as anhedral, mm-size interstitial grains. The latter display iron-rich rims that continue as films along olivine-olivine grain boundaries. The annealed microstructure and high iron-content of the dunites suggest significant thermal and chemical re-equilibration with basaltic magmas, consistent with their position as lenses outcropping in contact with metagabbros in the eastern border of the Balmuccia Massif (Boudier et al., 1984). Despite the similar annealed microstructures, dunite VS15 displays a much stronger olivine CPO than dunite VS14 (Fig. 1b,c).

## 4. Mechanical results

Despite the contrasting initial olivine grain size and CPO of the samples, the initial flow stresses show a limited range of variation, between 105 MPa and 165 MPa (Fig. 2). The fine-grained harzburgite 98OA87 displayed the highest flow stress and the sample cored parallel to the maximum concentration of [010] axes in coarse-grained dunite VS15 (VS15-b1) exhibited the lowest flow stress. Most coarse-grained samples started to flow between 110-150 MPa. However, pinpointing the yield stress for the coarse-grained dunites is difficult, since many samples, like VS14-4, VS14-7, VS14-9, VS15ab2, and VS15-a1, show continuous immediately post-yield hardening up to 10 or even 20% of bulk extension (Fig. 2).

The evolution of the mechanical behavior with increasing strain varies from sample to sample, reflecting differences in deformation behavior due to the contrasted initial microstructures and olivine CPO. For bulk strains > 15-20%, the differential stresses displayed in Fig. 2 are apparent ones. They were not corrected for the effects of heterogenous deformation, because the evolution of the shape of the sample over time cannot be assessed. The distribution of differential stress and strain along the length of the specimen (Fig. 3) may only be determined at the end of the test. The apparent strain weakening observed in the mechanical curves of samples 98OA87, VS15a1, VS15ab1, VS15ab1, and VS15b1 (Fig. 2) results therefore, partly or totally, from the cross-sectional area reduction. The continuous apparent weakening, starting at bulk strains as low as 7%, displayed by the fine-grained harzburgite 98OA87 is consistent with progressive reduction in the cross-sectional area due to strain localization in a wide necking zone. The coarse-grained samples VS15 only showed apparent weakening for bulk strains >20%, consistently with the homogeneous deformation of the coarse-grained samples in experiments stopped at strains > 20%. This apparent weakening probably results mainly from necking, but at the end of the experiments growth of fractures at high angle to the extension direction also contributed to it (Fig. 2). This process led to the ultimate failure of samples VS15-a1 and VS15-b1, which were deformed to >40% bulk strain. In contrast, sample VS14-9 displayed no apparent weakening despite clear necking and the mechanical data for sample VS14-7 recorded solely an abrupt decrease in stress after 40% bulk strain, which is probably associated with the growth of the fracture visible at the center of the cylinder in Figure 2. The absence or limited apparent weakening in the VS14 coarse-grained peridotites points to a hardening process, which counteracts the geometrical weakening due to necking in these experiments. The wide variability of hardening behaviors among the VS14 and VS15 samples in the initial stages of the experiments may be explained by the coarse grain sizes relative to the sample diameter. The mechanical behavior in each experiment may therefore depend on the orientation of the individual grains in a given section of the sample. The analysis of the microstructure of the deformed samples corroborates this hypothesis.

The final variation of axial differential stress, mean stress, and axial strain along the length of the 4 necked samples, which did not ultimately rupture, is presented in Figure 3. Necking resulted in up to three-fold enhancement of the differential stress compared to the specimen ends in all samples (Fig. 3). Maximum strains in the neck are more than five-fold the bulk strain. Focusing of strain in the neck implies a local increase of strain rate. If one considers that at the end of the experiment, strain is concentrated over <5 mm of the length of the sample, strain rates in the neck are $>5\times10^{-4}$ $s^{-1}$, that is, at least 5 times the initial ones. This combination of high strain rate, strain, and stress at the end of the test is reflected in the microstructure of the neck region in the recovered samples. Note that due to the non-linearity of the dislocation creep behavior, which in olivine is characterized by a stress exponent ~3 (cf. review in Hirth and Kohlsted 2003), a five-fold strain rate enhancement should result in an increase of the flow stress by a factor 1.7. Thus, the enhancement of the final stresses in the neck relative to the initial flow stresses by a factor 1.75 displayed by sample VS15-ab1, suggests that this sample displayed nearly steady-state flow up to large finite strains. The higher stress enhancement (factor ~3) determined for the other coarse-grained samples implies they were subjected to mild hardening. In contrast, for the fine-grained harzburgite, the final stresses in the neck are similar to the yield stress (~165 MPa) despite the increase in strain rate by up to a factor 5. This implies that the fine-grained harzburgite displayed significant strain-softening.

The peak values of axial differential stress in two coarse-grained samples exceed 300 MPa (the confining pressure) by up to 50 MPa of real tensile stress (Fig. 3). These tensile stresses are limited to the neck. High values of compressive load are maintained at the ends of the sample, preventing the separation of the sample ends from the loading pistons. These observations indicate that olivine-rich rocks deforming by viscoplastic processes at 1200°C can support true tensile stresses without fracturing. However, fractures started to develop in all necked coarse-grained samples and samples VS15-a1 and VS15-b1 ultimately failed at similar or even lower apparent differential stresses as that of sample VS14-9, suggesting that the conditions at the neck were close to the tensile limit at the temperatures and strain rates sampled in the experiments.

## 5. Microstructure evolution

In the coarse-grained dunites, olivine accommodates most of the deformation. It displays extensive evidence for deformation by dislocation creep, such as: elongation parallel to the stretching axis and strong undulose extinction, which evolves with increasing strain to closely-spaced planar subgrain boundaries, then to a network of irregularly-shaped to polygonal subgrains, and finally into recrystallized domains (Fig. 4). The importance of dislocation creep in olivine is corroborated by the analysis of the intracrystalline misorientation, which documents a steady increase in the geometrically-necessary dislocation (GND) density in olivine with increasing strain (Fig. 5). Enstatite and spinel usually preserve their original shapes. They may be locally elongated parallel to the stretching direction, but less so than olivine.

The strain distribution is highly heterogeneous. Within a sample, the intensity of deformation of olivine depends on the location (whether in the neck or outside, Fig. 4) and the orientation of the crystal relative to the imposed extension. The first-order control on the strain and stress distribution by the necking process is recorded by the spatial variation in the GND density in olivine (Figs. 6 and 7). Recrystallization restricted to the neck region also documents focusing of strain and stress. However, even coarse-grained samples deformed to < 20% bulk strain, which did not neck, display a heterogeneous deformation at the grain scale, recorded by the spatial variation of the GND density revealed by the KAM proxy in olivine (Figs. 5 and 6). Crystals initially well oriented for glide of dislocations of the [100](010) and [100](001) systems (bluish to purple grains in Figs. 7 and 8) are elongated and have a high density of subgrains, the boundaries of which are delineated by high KAM values (Figs. 5 and 6). In contrast, initially poorly-oriented crystals (greenish grains in Figs. 7 and 8) have more equant shapes and lower densities of GND, recorded by lower KAM values (Figs. 5 and 6).

Recrystallization occurs preferentially within grains well-oriented to deform by dislocation glide, often starting at the contacts with less well-oriented olivine crystals or with pyroxenes and extending towards the interior of the grain (Fig. 4). Recrystallized grains have polygonal shapes and average sizes ranging from 11 to 13 $\mu$m (Table 2). The detailed study of the recrystallization microstructures and the analysis of the effect of recrystallization on the CPO evolution, based on higher resolution EBSD maps of selected zones from the neck region of all samples deformed to > 20%, is presented in a companion article (Lopez-Sanchez et al., 2021).

Another characteristic microstructure of the coarse-grained peridotites is the development of heterogeneous intragranular deformation in grains poorly oriented for dislocation glide in the form of kink bands (Figs. 8, 9, 10). The kink bands have lens- or flame-like shapes and accommodate variable misorientations (from a few degrees to up to 85°) predominantly around the [001] axis (Fig. 10 and supplementary material Fig. S2), though rotations around <104>, <014>, <114>, <214>, and <142> were also documented (Supplementary material Fig. S2). In highly deformed domains, close to the neck, recrystallization develops along kink boundaries (Fig. 11).

Although only a single test was performed on the fine-grained harzburgite 98OA87, the evolution of the microstructure with increasing strain can be characterized by the comparison of EBSD maps from the necking region, where the sample was submitted to high stresses and strain rates, accumulating high finite strains, with those from the low strain zone close to the pistons, where the initial microstructure was largely preserved (Fig. 12). Since the starting material had already a partially recrystallized initial microstructure, the changes are less marked. The imposed extensional strain increases the elongation of the olivine porphyroclasts (which may attain aspect ratios as high as 10:1), creating a strong stretching lineation. In both low strain and high strain (neck) zones, the olivine porphyroclasts display a high density of subgrain boundaries and sinuous grain boundaries, but subgrain shapes change and the sizes of subgrains and the length scale of the sinuosity of the grain boundaries are smaller in the neck. In the low strain zone, straight subgrain boundaries normal to the grain elongation predominate, whereas olivine porphyroclasts in the neck have subgrain boundaries with more irregular shapes and variable orientations, which often form closed loops. Increasing extensional strain also produces a marked increase in the recrystallized fraction, which varies from ~20% in the low strain zone to 34% in the neck (Fig. 12). Dynamic recrystallization occurred mainly along grain boundaries by subgrain rotation and minor grain boundary bulging. It led to significant grain size reduction; average recrystallized grain sizes vary from 16 $\mu$m in the low strain zone, where they were probably inherited from the natural deformation forming the initial mylonitic microstructure, to 9 $\mu$m in the neck (Table 2). Effective dynamic recrystallization probably accounts for the maintenance of low mean intragranular misorientations in olivine even in the neck zone (Fig. 5). As in the coarse-grained peridotites, the deformation is essentially accommodated by olivine. Coarse pyroxene and spinel grains remain largely undeformed (cf. coarse pyroxene in the neck region in the photomicrograph in Fig. 2), but may be locally stretched and fractured at high angle to the imposed extension (Fig. 12d,e).

## 6. Olivine Crystal Preferred Orientations

Olivine in the starting material for the fine-grained peridotite 98OA87 sample had already a CPO, characterized by a poor alignment of [100] at low angle to the preexisting stretching lineation (Fig. 1a). The extension was applied roughly parallel to this direction. Comparison of this initial olivine CPO to that in the low strain domain close to the piston and to that in the necking region (Fig. 12c,f) highlights: (i) a marked enhancement in the strength of the CPO and (ii) reorientation of the olivine crystals with development of a strong maximum of the <101> axes parallel to the stretching direction.

The coarse grain size in the other samples hinders the evaluation of the CPO evolution with increasing strain, as either finite strains were low (bulk strains <20%) or they were heterogeneous along the sample so that representative volumes for each strain intensity could not be analyzed. However, the increase in the area filled by blueish to purplish grains, which have their [100] to <101> axes aligned with the imposed extension as the bulk strain of the sample is increased and within the neck zone in the necked samples, suggests a similar tendency in the CPO evolution (Figs. 8, 9).

**7. Water content in olivine and partial melting**

Unpolarized FTIR spectra acquired on olivine and opx grains in the three starting materials (98OA87, VS14, VS15), in one annealed sample of the coarse-grained dunite VS14 (VS14An), and in two coarse-grained dunites deformed to low bulk strains (VS14-3, VS15b3) are displayed in Figure 13. In all analyzed olivine grains, the classic OH bands located at 3572 cm$^{-1}$, 3525 cm$^{-1}$ and 3240 cm$^{-1}$ are present, but are barely visible (cf. for comparison Demouchy and Bolfan-Casanova, 2016). They represent very small amount of hydrogen in the olivine structure ($\leq$ 2 ppm H$_2$O wt., Table 3). It is important to note that there is no marked different between the starting material and the annealed or deformed VS14 and VS15 samples. All have almost completely dry olivine.

Olivine grains in the analyzed slab of sample 98OA87 were too small. Thus grain boundaries, which systematically contain minute amounts of serpentine (OH band at 3690 cm$^{-1}$), could not be avoided in the analyses. However, orthopyroxene porphyroclasts were large enough for the acquisition of spectra free of contamination by hydrous minerals. These orthopyroxene porphyroclasts display the typical OH bands at 3564 cm$^{-1}$, 3520 cm$^{-}$

$^{-1}$ and 3420 cm$^{-1}$ and contain 76 ppm H$_2$O wt. If equilibrium is assumed, such a water content in orthopyroxene would correspond to 15 ppm H$_2$O wt. in olivine based on a partition coefficient D$_{opx/ol}$ of 5 (Hirth and Kohlstedt, 1996, Demouchy et al., 2017).

Small amounts of glassy material, indicating partial melting, were documented in all deformed samples, including the fine-grained harzburgite (Supplementary Material Fig. S1). A precise quantification of the melt fraction was not possible, as the analyses were limited to the 2D sections exposed by cutting the samples parallel to the extension axis. However, glassy material occurs in very small amounts (always <1% of the observed surface). Moreover, it always occurs in isolated pockets at triple junctions or along grain boundaries, in most cases in the vicinity of pyroxenes and spinel. These pockets are usually <300 $\mu$m wide and are composed of an association of small polygonal grains of olivine and glassy material in variable proportions. Larger pools of glassy material (up to 800 $\mu$m long and <100 $\mu$m wide) with feather-like structures were only observed along a few grain boundaries at a high angle to the extension direction in samples VS14-8 and VS15-ab2. Evidence for concentration of glassy material in the neck regions, indicating melt migration controlled by variations in effective mean pressure along the sample, was not observed. Most recrystallized domains in the neck are glass-free at the scale of the optical and scanning electron microscope observations. These observations are consistent with partial melting controlled by dehydration of alteration phases. These alteration phases were in most cases not visible optically in the starting material, but their presence along grain boundaries was clearly detected by FTIR (Fig. 13).

## 8. Discussion

*8.1. Limitations and advantages of the present experimental data*

Owing to the coarse initial grain size of the VS14 and VS15 dunites (Fig. 1), the experimental samples of these rocks are not representative volumes. The experimental cylinders have on average 2 to 5 grains across the initial diameter (Figs. 8, 9). By consequence, the mechanical behavior of these samples is strongly dependent on the behavior of individual grains in any transect and cannot be described by homogenization schemes. Thus, strain localization is not only controlled by the non-linearity of the stress-strain relationship (Hart, 1967; Rutter, 1998), but principally by the location of grains in easy glide orientations. This explains the high variability of mechanical behavior among the coarse-grained samples (Fig. 2). This effect is enhanced by the presence of an initial CPO, as in VS15 (compare Figs. 8 and 9). However, the exceptionally coarse initial grain size of olivine in these samples (>1 mm), compared to the dominantly <50 $\mu$m grain sizes used in deformation experiments on sintered fined-grained olivine polycrystals (e.g., Bystricky et al., 2000; Faul et al., 2011; Hansen et al., 2016, 2014, 2012; Karato et al., 1986; Thieme et al., 2018; Zhao et al., 2009), allows investigation of processes that are usually not observed in experiments but may occur in nature, where millimetric size olivine dominates, such as the role of the strong viscoplastic anisotropy of olivine crystals and interactions between crystals with different orientations on dynamic recrystallization and development of kinks (Figs. 4, 10, 11). Moreover, the comparison between the mechanical and microstructural data of the two coarse-grained samples to that of the fine-grained one permits investigation of the links between mechanical response and microstructure evolution over a range of grain sizes akin to that observed in nature.

*8.2. Initial strength*

Analysis of the mechanical data shows that the fine-grained harzburgite displays the highest yield stress – 165±5 MPa (Fig. 2). These stresses are lower than that predicted for the present deformation conditions based on experimental data from dry synthetic aggregates of San Carlos olivine (Fo90) deformed by dislocation creep in axial compression (≥300 MPa, Hirth and Kohlstedt, 2003; Thieme et al., 2018). They are, however, within the range of those measured for natural dunites without pre-deformation thermal treatment at similar temperatures and strain rates: 90 and 175 MPa for samples with mean initial grain sizes of 100 and 900 $\mu$m, respectively (Chopra and Paterson, 1981). In these experiments, similarly to the present ones, the peridotites deformed in the presence of minor melt fractions due to breakdown of hydrous phases present along grain boundaries in the starting material.

The dunites yielded at even lower differential stresses, between 115 and 150 MPa. The higher iron contents of olivine (Fo82-83, compared to Fo90 in the harzburgite) may explain the lower initial strength of the dunites. Based on the flow law by Zhao et al. (2009) and considering solely the influence of the variation in iron content of olivines, the dunites should support stresses on the order of 0.74 times that of the harzburgite, that is ~122 MPa, which is in the lower part of the range of the observed values. However, given the large difference in initial average grain sizes, one could expect a higher contribution of diffusional or grain boundary processes to deformation in the fine-grained harzburgite

(area-weighted mean grain size of 170 $\mu$m), which should reduce its strength relative to that of the coarse-grained dunites. Considering a grain size exponent of -2 and a stress exponent of 3 (Hirth and Kohlstedt, 2003), an increase in average grain size from 170 $\mu$m to 1 mm should result in an increase in strength by a factor 12. If one considers that samples with average grain sizes ≥250 $\mu$m deform in a grain-size independent regime the difference reduces to a factor 1.3. Combining the effects of composition and grain size variations, the coarse-grained dunite samples are expected to display yield stresses similar or higher than the fine-grained harzbugite sample. However, all dunites yielded at lower differential stresses than the harzburgite. This implies that either grain size-related weakening played a minor role in the present experiments or that other processes further weakened the dunite samples. The present data also contrast with previous results on natural dunites, which showed lower strengths for the finer-grained samples (Chopra and Paterson, 1981). The lower strength of the coarse-grained dunites relative to the fine-grained harzburgite also cannot be explained by water weakening, since the hydrogen concentrations in olivine are lower in the dunites (Fig. 13 and Table 3). A possible explanation could be the presence of higher melt fractions in the dunite samples. However, all initial materials had minor contents of hydrous phases along grain boundaries, which dehydrated during the experiments to produce localized melt pockets. Glassy material was indeed observed in all deformed samples. However, this material represents <1% of the analyzed sections and is not interconnected (Supplementary Material Fig. S1). Such low non-interconnected melt fractions should not significantly lower the bulk strength of a peridotite (Kohlstedt and Chopra, 1994; Hirth and Kohlstedt, 2003).

As discussed in the previous section the extreme variability in mechanical behavior among the dunites stems from the combined effect of very coarse grain sizes relative to the sample diameter and of the high viscoplastic anisotropy of olivine crystals. Indeed, the strength of an olivine crystal perfectly oriented to deform by dislocation glide on the easy [100](010) system is at least three times lower than that of a crystal perfectly oriented to deform by dislocation glide on the hard [001](010) system (Bai et al., 1991) and a crystal oriented with its [100] or [010] axes parallel to the imposed extension theoretically cannot deform by dislocation glide, because resolved shear stresses on all possible slip systems would be null. The observed initial strengths of the VS15 samples subjected to axial extension in different directions relative to the olivine CPO (Fig. 2) are consistent with the relative strengths predicted by viscoplastic self-consistent simulations using the same CPO and boundary conditions. The exception is sample VS15-b1, which should have presented a higher yield strength, similar to that of samples VS15-a1 and VS15-ab1 (cf. Supplementary Material). Analysis of the olivine orientation map for VS15-b1 (Fig. 9) shows, nevertheless, that deformation in this sample was concentrated in a zone where most crystals had their [100] and [010] axes oblique to the imposed extension and hence that displayed a lower strength than the remainder of the sample, further corroborating the major effect of viscoplastic anisotropy on the mechanical behavior of these samples.

*8.3. Strength and microstructure evolution with increasing strain*

The fine-grained harzburgite and the coarse-grained dunites displayed markedly different strength evolutions (Fig. 2), which are related to different strain distributions at the macroscopic scale, as illustrated by the analysis of the final shapes of the samples (Fig. 2 and 3) and their different microstructures (Figs. 4 to 12).

The fine-grained harzburgite showed a well-defined yield point at ~2% strain, consistent with a sharp transition from elastic to viscoplastic deformation, and a short duration steady-state followed, from 7% of bulk strain onwards, by a steady decrease in apparent strength. This decrease in apparent strength may be largely explained by the necking of the sample, with the deformation progressively focused in smaller and smaller volumes; the progression of necking leading to continuous decrease of the cross-section of the actively deforming region (Figs. 2 and 3). Analysis of the local differential stresses and finite strain attained at the end of the test shows that the maximum stresses at the neck (169±15 MPa) are similar to the yield stress of this sample (167±5MPa) despite local strain rates at least 5 times higher (finite strains in the neck are up to five times the bulk strain at the end of the experiment). If this sample deformed dominantly by dislocation creep, with a stress-dependence to a power of 3, an enhancement in strain rate by a factor 5 should result in a stress enhancement by a factor 1.7. The present data imply that the microstructural evolution in this sample, in particular the effective dynamic recrystallization, counteracts this strain-rate hardening. This inference is corroborated by the low area-weighted mean GOS of olivine in the neck zone (Fig. 5b). Effective grain size reduction by dynamic recrystallization – the recrystallized volume increases from ~20% to 34% and the average recrystallized grain size decreases from 16 $\mu$m to 9 $\mu$m from the low strain zone to the neck (Fig. 12) – may also increase the contribution of diffusional processes to deformation in the neck. Yet, in itself, the marked reduction in grain size indicates that dislocation creep played a major role during most of the deformation of the sample.

Another possible source of strain weakening during deformation by dislocation creep is the evolution of the CPO. To isolate this effect, we ran viscoplastic self-consistent simulations (VPSC) where we compared the strength of aggregates with the olivine CPO measured in the low strain zone and in the neck region of sample 98OA87 (Fig. 12) subjected to boundary conditions reproducing those of the experiments, using a similar approach to Mameri et al. (2019). Input parameters and full results of the VPSC simulations are presented in the Supplementary Material. These simulations predict that the moderate change in the CPO between the low strain and the neck region would result in moderate hardening, by a factor 1.1, if the sample was deformed solely by dislocation creep. This hardening stems from to the progressive rotation of the olivine [100] axes towards the extension direction, which results in a progressive decrease of the resolved shear stresses on the easy [100]{0kl} systems. However, the VPSC simulations also predict that in absence of dynamic recrystallization the CPO evolution would have been faster and the associated hardening greater at >100% of finite strain attained in the neck, stronger. In conclusion, dynamic recrystallization weakened this sample via two processes. It kept intragranular dislocation densities low (Fig. 5) and changed the CPO evolution, reducing the hardening associated with the CPO evolution produced by deformation in axial extension by dislocation glide alone (cf. Supplementary Material Fig. S3).

Surprisingly, the differential stress predicted based on the average recrystallized grain size in the neck using the paleopiezometric relation of Van der Wal et al. (1993) – 219 MPa (Table 2) – is significantly higher than that estimated from the macroscopic data

- 165±15 MPa. We do not have a simple explanation for this discrepancy, as this piezometer has been established based on experiences on similar starting materials and deformation conditions. However, Van der Wal et al. (1993) already observed that, at the same stress level, recrystallized grain sizes in samples of the initially fine-grained Anita Bay dunite (100 $\mu$m) were systematically smaller than those of the initially coarse-grained Aheim dunite samples (900 $\mu$m). The piezometer fit, which is based on the entire dataset, overestimates therefore the stresses for the Anita Bay samples (cf. Fig. 2 of Van der Wal et al., 1993).

In contrast, the coarse-grained dunites show an extended hardening behavior up to 10-20% bulk strain and either a weak apparent softening or an apparent steady-state behavior despite necking at >20% bulk strain (Fig. 2). This behavior implies continuous hardening of the samples, though the apparent softening observed for the necked VS15 samples suggest a decreasing rate of hardening at high strains. The observed mechanical behavior is consistent with the measured steady increase in the mean GOS with increasing strain (Fig. 5), that records a steady increase in the dislocation density. As previously discussed, the differences in the mechanical behavior between the coarse-grained samples result from differences in the orientation of the olivine crystals that comprise each sample, with clear concentration of strain in those crystals in easy glide (soft) orientations. In all samples, necking occurred in domains with a high concentration of such crystals (purple crystals in Figs. 8 and 9). Concentration of the deformation in domains with a higher volume of olivine grains in soft orientations probably resulted in macroscopic softening of the sample, but the latter was counteracted by the local increase in strain rate (to compensate for deformation accommodated in the smaller volume of the neck) and by the increase in the intragranular dislocation density, which is markedly higher in the neck zones (Figs. 6 and 7).

Flow stresses at the neck of the coarse-grained samples estimated based on the minimum final diameter (ignoring the fractures) range between 325 and 470 MPa (Fig. 3). The lowest enhancement of the final stresses in the neck relative to the initial flow stresses (by a factor 1.75) is displayed by sample VS15-ab2. This value is equivalent to the increase in stress associated with the increase in strain rate in the neck estimated from the finite strain distribution, suggesting that this sample attained nearly steady-state flow. In all the remaining coarse-grained samples, the final stress in the neck was higher by a factor three than the initial flow stress, corroborating the inference of continued microstructural hardening in addition to the increase in stress associated with the increase in strain rate. This is despite the onset of dynamic recrystallization in all necked samples. Recrystallized volumes remained nevertheless small and much of the strain was still accommodated by dislocation glide and kinking in the coarse olivine grains (Fig. 4).

As a consequence of this hardening, the peak values of axial differential stress in the neck of the coarse-grained samples VS14-9 and VS15-ab1 exceeded the confining pressure, leading to up to 50 MPa of real tensile stress accommodated by ductile flow (Fig. 3). Ductile metals and ceramics can support large true tensile stresses without cavitation or fracture, as solid-state diffusion may suppress the growth of crack-like openings along grain boundaries. Olivine aggregates can do the same at high temperature, although separation and ultimate failure are expected ultimately to supervene. Up to five-fold elongation was obtained in initially extremely fine-grained (< 500nm) aggregates of forsterite + minor diopside or periclase in tensile tests at 1450°C and ambient pressure, before failure at stresses of ~30 MPa (Hiraga et al., 2010). In the present experiments, the development of a grain shape fabric probably delayed fracturing as it reduced the area of grain boundaries subjected to interface-normal extension (Fig. 4a). However, fractures eventually started to develop in all necked samples and samples VS15-a1 and VS15-b1 ultimately ruptured at similar or even lower apparent differential stresses as that of sample VS14-9. This suggests that 50 MPa is close to the tensile limit for olivine rich-rocks at the tested grain sizes, temperature and pressure conditions.

The peak values of axial differential stress in the neck of the coarse-grained samples estimated from the macroscopic data (250 to 365±15 MPa) are significantly higher than those estimated based on the recrystallized grain sizes (159 to 182±5 MPa). This suggests a delay in the reequilibration of the microstructure, with the recrystallized grain sizes recording the lower differential stresses that preceded the final stress at which the test was terminated. Recrystallized grain sizes in the coarse-grained dunites are larger than those determined for the fine-grained harzburgite (Table 2), despite the much lower maximum stress at the neck attained in the fine-grained harzburgite (Fig. 3). This apparently contradictory observation might be explained by a more progressive evolution and more homogeneous strain and stress conditions in the fine-grained sample, allowing for a more effective re-equilibration of the microstructure.

Comparison of the microstructures of the highly strained coarse-grained dunites and the fine-grained harzburgite reflects the higher volume fraction of newly recrystallized grains in the latter. The mechanical data point to effective strain softening in the fine-grained harzburgite, which results in necking almost immediately after yielding, whereas the coarse-grained samples harden or at best attain steady-state flow conditions. Altogether, these observations are consistent with the major role of dynamic recovery and dynamic recrystallization in counteracting microstructural hardening (Nicolas and Poirier, 1976; Rollett et al., 2004). However, why was dynamic recrystallization more effective in the fine-grained harzburgite? The present data hint at a major role of the higher initial density of grain boundaries, which acted as preferential nucleation sites in the fine-grained harzburgite. In the coarse-grained dunites, a large part of the recrystallization process is intragranular, requiring significant densities of dislocations of at least three independent slip systems to create closed subgrains that may evolve into recrystallized grains. By consequence, dynamic recrystallization was only effective in the neck zone and occurred preferentially in soft grains close to the contact with hard ones, implying that it required both high strain and high stresses, which are only achieved locally, due to intergranular interactions. We discuss this point in a companion article (Lopez-Sanchez et al, 2021), which is based on the analysis of the higher resolution EBSD maps of the recrystallized domains.

*8.4. Dislocation creep and kink bands*

In both the coarse- and fine-grained samples, the microstructures and CPO evolution point to deformation mainly accommodated by dislocation creep of olivine, with activation of multiple slip systems, but predominance of [100] glide (Figs. 4, 8, 9, 12), accompanied by dynamic recrystallization in highly strained zones (Figs. 4, 11, 12). However, the coarse-grained dunites also show microstructures that are rarely observed in olivine: kink bands (Fig. 10, 11, and Supplementary Material Fig. S2). Kink bands are planar deformation features characteristic of highly anisotropic materials, which possess insufficient slip systems to satisfy the Taylor/Von Mises criterion for homogeneous plastic flow at constant volume (Barsoum et al., 1999; Burnley et al., 2013; Nicolas and Poirier, 1976; Raleigh, 1968). They provide an additional degree of freedom for deformation by producing sharp reorientation of volumes of crystals that are poorly oriented to deform by dislocation glide on the existing systems. This is consistent with observation of kink bands in olivine crystals with the [010] axis subparallel to the imposed extension (Figs. 8, 9, 10 and Supplementary Material Fig. S2). In the present experiments, misorientations across kink bands are essentially accommodated by rotations around [001], suggesting that dislocations of the dominant [100](010) slip system played an important role in their formation. The onset of recrystallization along kink boundaries (Fig. 11) is consistent with local stress concentration and high dislocation densities. However, misorientation angles across kink boundaries are too high to be explained solely by the accumulation of dislocations, suggesting that microfracturing may also play a role. Cracks assisting strain accommodation in kink bands were indeed observed by TEM in highly anisotropic ceramics, in particular for high misorientations (Barsoum et al., 1999).

Kink bands in olivine are extremely rare in naturally deformed peridotites, but have been documented in meter to centimeter-scale shear zones in the Finero massif in the Alps (Matysiak and Trepmann, 2015) and in the ductile root of detachment faults in the South West Indian Ridge (Bickert et al., 2021). In both cases, initial olivine grain sizes were millimetric to centimetric and the development of kinks was followed by dynamic recrystallization leading to extreme grain size reduction. Kinks were also observed in $Mg_2GeO_4$ olivine deformed experimentally under high-pressure (0.6-1.3 GPa) and high-stress conditions (1.3-2.5 GPa; Burnley et al., 2013). The present observations and these three examples share many characteristics: (1) kinks limited to crystals in hard orientations, (2) highly variable misorientations along the kink boundary, which may attain 90°, (3) consistent crystallographically-controlled rotation axis across the kink bands, which point to a contribution of dislocations of the dominant slip system, (4) high differential stresses, >100 MPa for olivine, and (5) coarse grain sizes (relative to the sample dimensions in the experiments), leading to dominant intragranular deformation by dislocation glide.

## 9. Conclusion

Natural upper mantle peridotites with contrasting initial microstructures and variable CPO subjected to axial extension at 1200°C and 300 MPa confining pressure showed similar initial strengths, ranging between 115±5 and 165±5 MPa, with the finer-grained sample displaying the highest yield stress. The lower initial strength of the coarse-grained samples may be partially explained by the higher iron content of olivine. However, the difference in composition did not suffice to counteract the expected contrast in strength due to the difference in initial grain size. Olivine in the coarse-grained dunites is almost totally hydrogen-free, so their lower strength cannot be explained by water weakening processes. All samples had minor contents of hydrous phases along grain boundaries, which triggered partial melting during the experiments. However, there is no clear evidence for higher melt fractions in the coarse-grained dunites. In all samples, glassy material composes <1% of the samples' area and is not interconnected. Altogether these observations imply that despite the fact that ca. 70% of the harzburgite is composed of olivine grains with 10 to 300 μm in diameter (median at 14 $\mu$m), grain-size sensitive processes did not significantly reduce its yield strength.

All samples deformed to > 20% bulk strain became necked. However, fine- and coarse-grained peridotites showed markedly different evolutions of both mechanical behavior and microstructure. The fine-grained harzburgite displayed, after yielding, a progressive decrease in the apparent differential stress consistent with the observed necking. It showed an increase in the olivine CPO intensity and recrystallized fraction and a decrease in the recrystallized grain size. These features recorded strain, strain rate, and stress focusing in the neck. The marked decrease in recrystallized grain size points to dominant dislocation creep despite the fine initial grain sizes. Stresses in the neck at the end of the experiment were similar to the initial yield stress despite significant strain localization, which implied strain rates in the neck up to five times higher than the initial ones. This indicates that this sample was subjected to significant strain softening. We infer that this softening is due to continued dynamic recrystallization, which kept dislocation densities low, favored the development of crystal orientations favorable for dislocation glide and, probably, enhanced the contribution of grain-size dependent creep processes.

In contrast, most coarse-grained samples showed hardening in the initial stages of deformation and no or little decrease of the apparent differential stress after necking. This is consistent with the increase in the bulk intragranular misorientation with increasing strain in these samples, absence of evidence for dynamic recrystallization in the non-necked samples, and limited recrystallized volumes in the necked ones. The coarse-grained samples systematically displayed a highly heterogeneous deformation. Crystals well oriented to deform by dislocation glide became elongated with marked undulose extinction, whereas those in hard orientations remained weakly deformed. The initiation of the neck was favored by local concentrations of crystals in soft orientations. In the neck, stress concentrations resulted in formation of kinks in crystals in "hard" orientations for dislocation glide, dynamic recrystallization in crystals in "soft" orientations, and, at bulk strains >40%, development of extensional fractures that produced a rapid drop in the apparent differential stress. Differential stresses in the neck estimated based on recrystallized grain sizes were significantly lower than those estimated using the neck minimum diameter, suggesting that the recrystallized grain sizes were established at the lower differential stress that preceded the final stress peak.

Differential stresses greater than the confining pressure, implying development of local true tensile stresses of ca. -50 MPa, were recorded in the neck regions of the coarse-grained dunite samples. This indicates that, at 1200 °C, olivine-rich rocks can sustain substantial tensile stresses before being subjected to rupture by fracturing. The highest recorded stress values are probably close to the ductile tensile limit of these samples under the present experimental conditions, as indicated by the eventual onset of fracturing and ultimate failure of some samples.

The take-home message of the present study is that variations in grain size representative of those observed in natural peridotites may result in significantly different evolutions of the mechanical behavior with increasing strain, even if the initial strengths are similar. Comparison of the mechanical and microstructural observations on the coarse and fine-grained samples suggests that the inhibition of dynamic recrystallization, probably due to the lower density of grain boundaries acting as nucleation sites, resulted in significant hardening and activation of high-stress deformation processes, such as kinking and fracturing, in the coarse-grained peridotites. This effect is particularly marked in the present experiments, where dynamic recrystallization is dominated by nucleation, because experimental strain rates largely overwhelm grain boundary migration rates. In nature, at much slower strain rates, the importance of nucleation in controlling the effectiveness of dynamic recrystallization to produce strain softening might be lower.

## Acknowledgments

We dedicate this article to A. Nicolas, an exceptional researcher who set the basis of the petrophysical analysis of mantle deformation. We are grateful for constructive comments from Greg Hirth and an anonymous referee. Senior experimental officer Robert Holloway (Rock Deformation Laboratory) provided invaluable training to WBI and maintained the equipment. Françoise Boudier and Adolphe Nicolas are thanked for the providing the Oman sample and David Mainprice for helpful discussions. EBSD analyses were performed at the SEM-EBSD CNRS-INSU national facility at Géosciences Montpellier. The IR-Raman technological Platform of University of Montpellier is acknowledged for the vibrational experiments. The experiments were carried out with support from an EU-Marie Slodowska-Curie postdoctoral fellowship to WBI. The data analysis was supported by the European Research Council (ERC) under the European Union's Horizon 2020 research and innovation programme (grant agreement No 882450 – ERC RhEoVOLUTION) and by a postdoctoral fellowship co-funded by the European Union and the Government of the Principality of Asturias (Spain) [grant number ACA17-32] within the FP7 Marie Skłodowska-Curie COFUND Action to MALS.

**Figures and Tables**

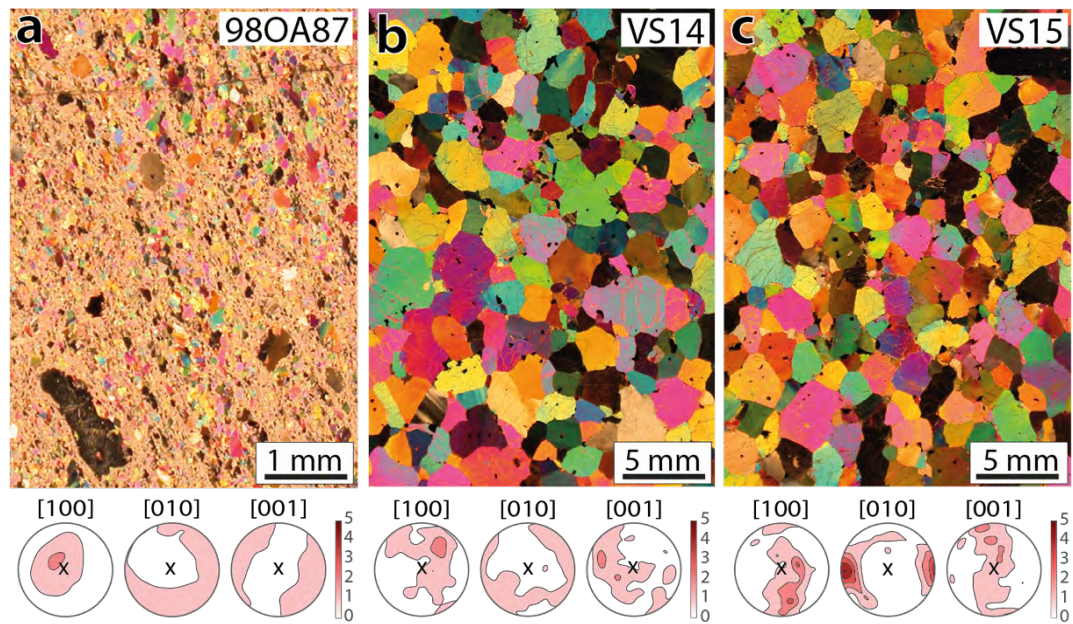

**Fig. 1.** Microstructure and olivine crystal preferred orientation (CPO) of the three starting materials: plane-polarized photomicrographs and pole figures representing the olivine crystals orientation relative to the long axis of the photo (x). In (a) the preexisting stretching lineation of the sample is at ~20° to this direction. In (c) the concentration of [010] is parallel to the direction of flattening of the olivine grains (short axis of the photo). Same color scale is used in all pole figures.

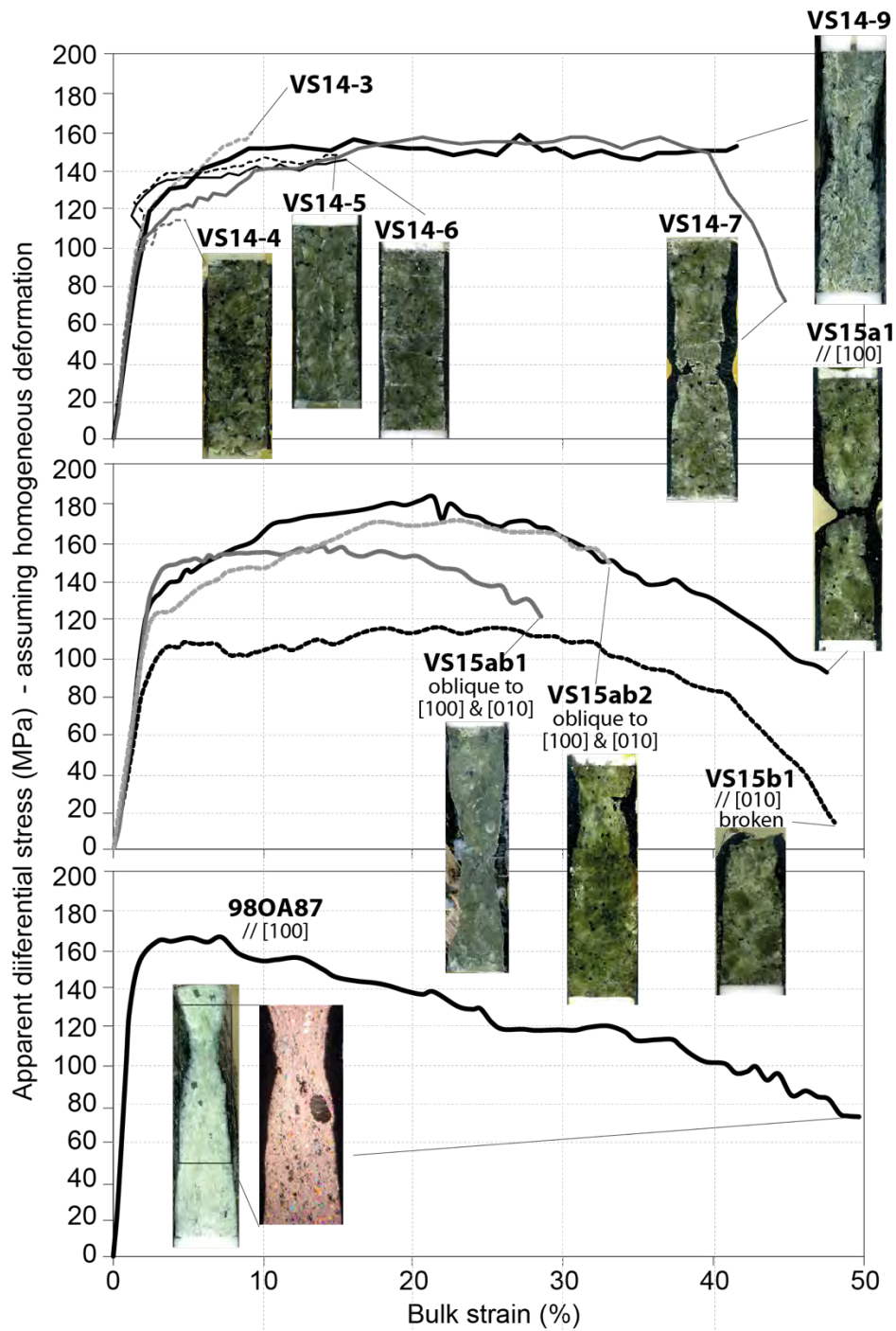

**Fig. 2.** Mechanical data for all successful experiments (Table 1). Apparent differential stresses (corrected for the strength of the iron jacket) are calculated assuming homogeneous reduction of specimen cross sectional area at constant volume along the length of the sample. They represent the actual differential stresses for samples that did not neck. In the samples subjected to necking, stresses vary strongly along the sample length (Fig. 3). Strains are bulk strains, that is, the overall change in length (after correction for axial apparatus distortion) divided by the original sample length. Final sample shapes are illustrated by scans of the recovered samples cut parallel to the extension axis as close to the cylinder center as possible. For the fine-grained harzburgite 90OA87, a crossed-polars micrograph of the most deformed domain is also displayed; note the strong inclusion behavior of the coarse orthopyroxene crystal (dark rounded grain) close to the neck region. Orientations below sample names indicate the orientation of the core relative to the olivine CPO of the initial material. The different cores of dunite VS14 have random orientations relative to the very weak CPO of the starting material.

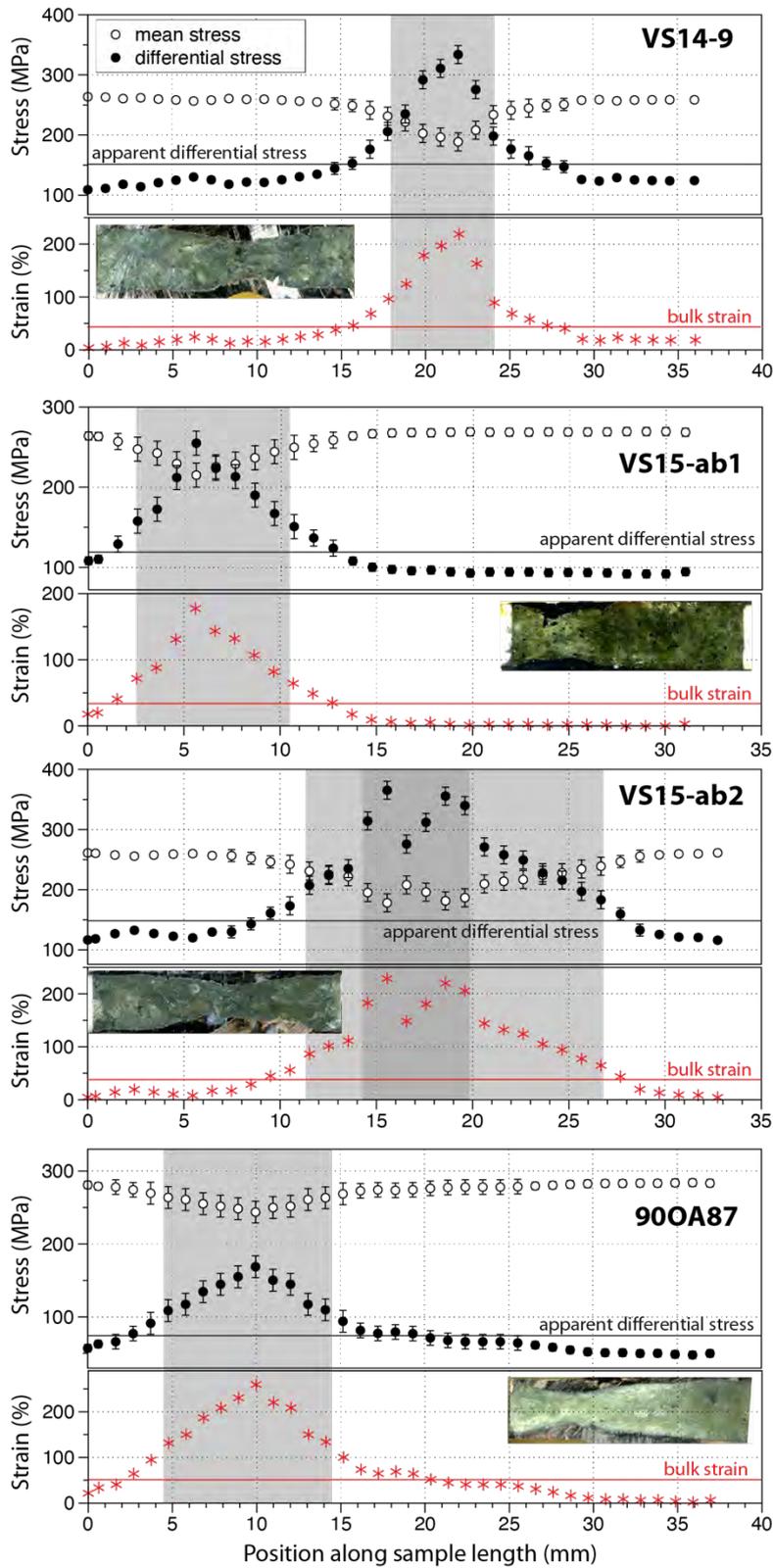

**Fig. 3.** Axial differential stress, mean stress, and extensional strain distributions along the sample length for the four necked samples that did not break. All three starting materials are represented. Apparent axial differential stress and bulk strain are indicated for comparison. The neck region, outlined in gray, is subjected to significantly higher stresses and strains; its width varies strongly from sample to sample. Sample VS15-ab2 shows abrupt variations in strain and stress, suggesting development of a new instability within the initial neck zone.

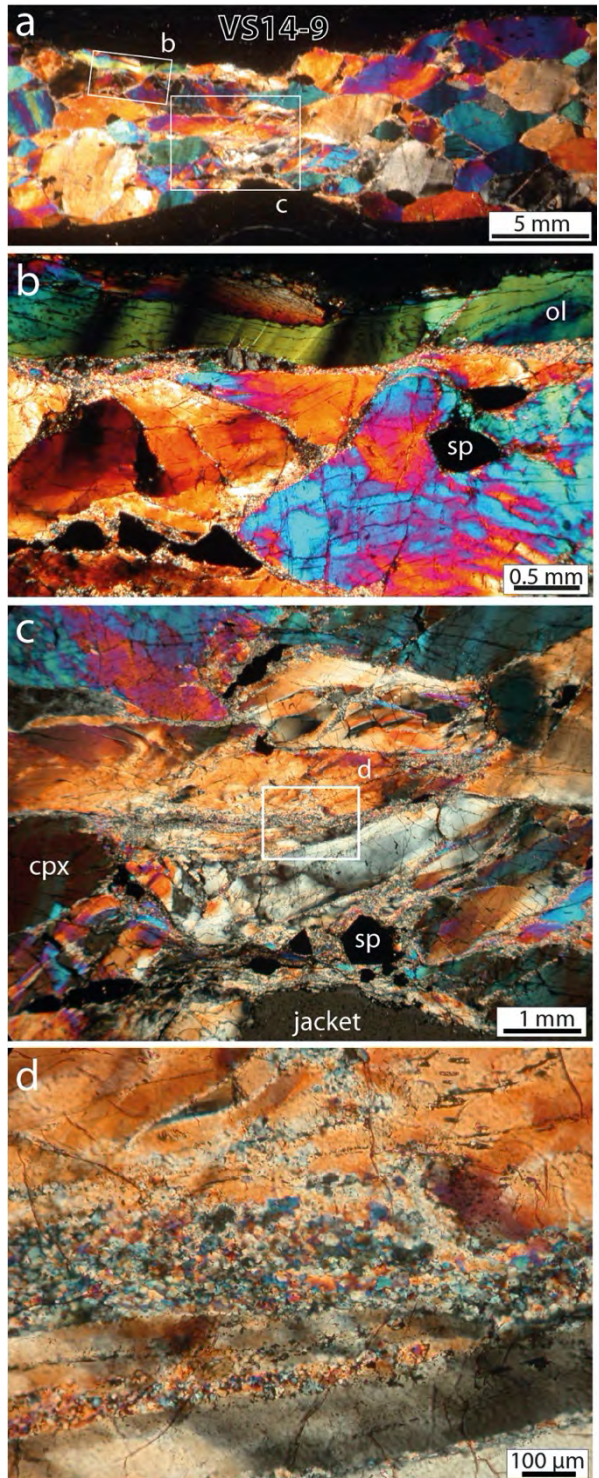

**Fig. 4.** Representative olivine deformation microstructures (coarse-grained dunite VS14-9). (a) Low enhancement cross-polars photomicrograph highlighting the contrast in microstructure due to strain concentration in the neck. (b) Detail of (a) illustrating an extremely elongated (aspect ratio ~ 20:1) olivine crystal (ol) with strong undulose extinction, a complex subgrain pattern and recrystallization seams in the neighboring olivine crystals, and coexistence between spinels (sp) undeformed and stretched by fracturing. (c) Detail of the neck region showing highly deformed olivine crystals crosscut by fine-grained recrystallization seams, displayed in detail in (d). The extension direction is parallel to the long side of the image in (a,c,d) and at ~10° to it in (b).

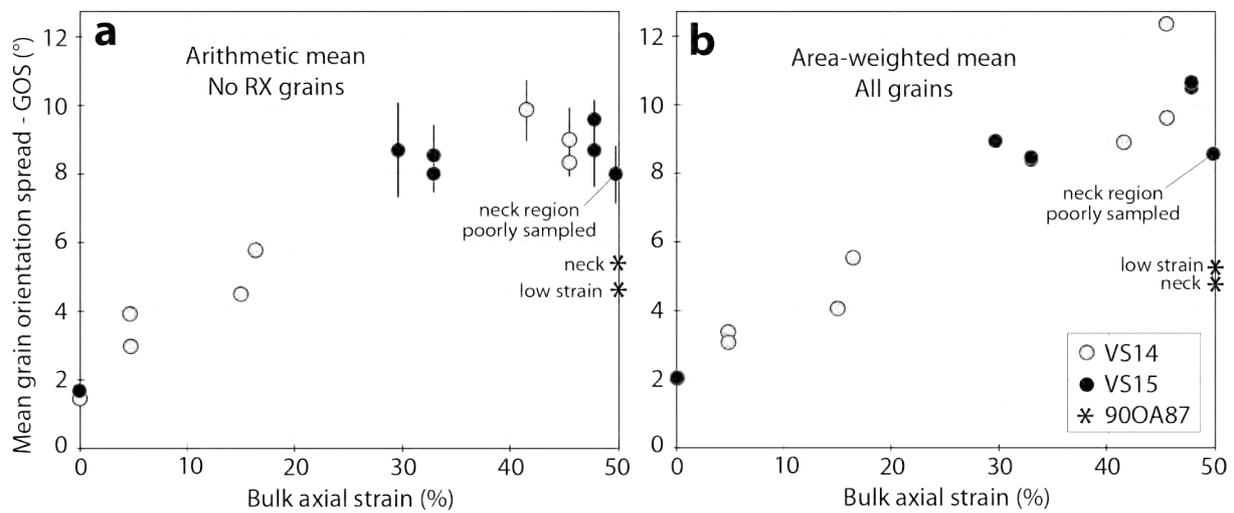

**Fig. 5.** Bulk (sample scale) intracrystalline misorientation in olivine represented as the grain orientation spread (GOS) as a function of the final bulk axial strain. (a) Arithmetic mean GOS excluding the recrystallized grains from the calculation; error bars indicate the standard error at 95% of confidence. RX means recrystallized grains. (b) Area-weighted mean GOS including all grains, which should be indicative of the contribution of the intragranular dislocation density to the sample strength. The lower area-weighted GOS relative to the arithmetic mean GOS in the neck region of sample 90OA87 (star symbols) reflects the contribution of dynamic recrystallization towards decreasing the intragranular dislocation density. The coarse-grained samples have similar or higher area-weighted GOS relative to the arithmetic mean GOS, consistent with their low recrystallized volumes. The higher area-weighted GOS values result from the presence in the mapped area of coarse grains with high intragranular misorientations. Data are presented in Supplementary Material Table S1.

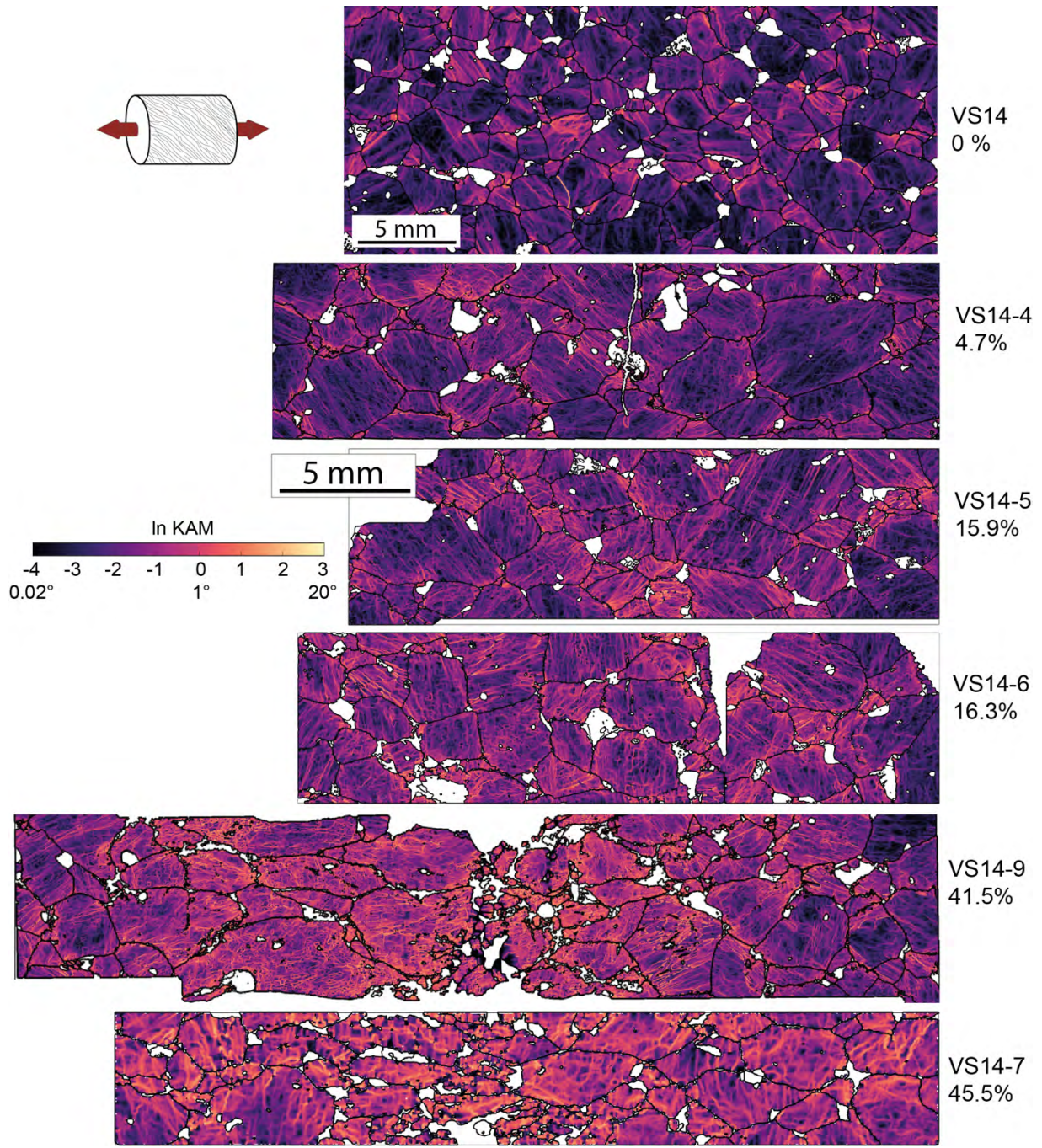

**Fig. 6.** Evolution of the geometrically-necessary dislocation (GND) density of olivine with increasing strain illustrated by KAM maps for VS14 dunites deformed to different bulk finite strains (numbers below sample name). High KAM values (hotter colors) delineate the subgrain structure. Grain boundaries are displayed in black. Pyroxenes, spinels, dynamically recrystallized olivine grains, and missing material (lost during preparation of the sections for EBSD analyses) are displayed in white. Scale is the same for all deformed sample maps. Note: (1) the on-average higher GND densities denoting stress and strain concentration in the neck domains of samples stretched by >20% bulk strain and (2) the heterogeneity in the GND spatial distribution in all samples denoting stress and strain heterogeneity at the grain scale.

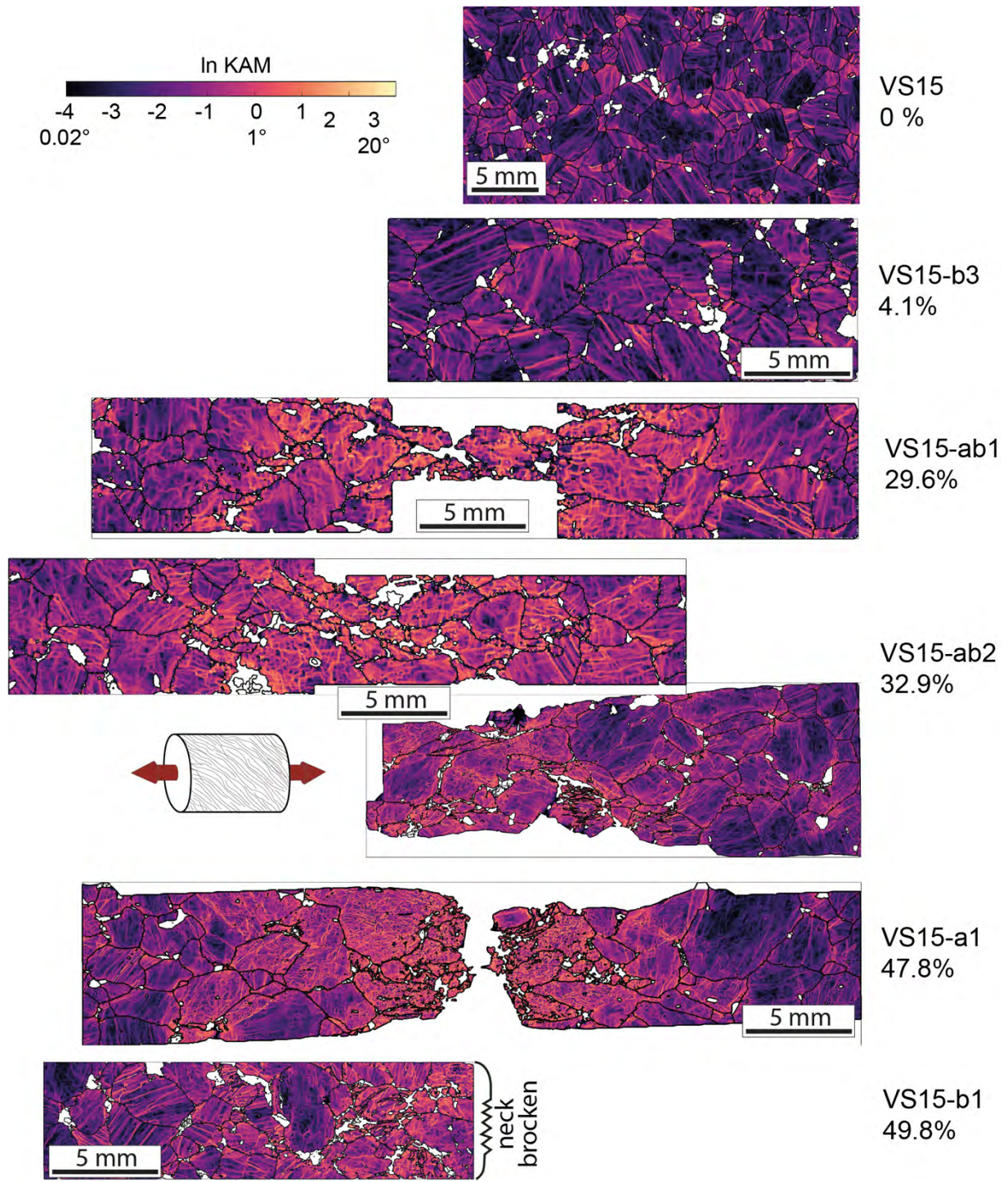

**Fig. 7.** Evolution of the geometrically-necessary dislocation (GND) density of olivine with increasing strain illustrated by KAM maps for VS15 dunites deformed to different bulk finite strains (numbers below sample name). High KAM values delineate the subgrain structure. Grain boundaries are displayed in black. Pyroxenes, spinels, dynamically recrystallized olivine grains, and missing material (lost during preparation of the sections for EBSD analyses) are displayed in white. As in Fig. 6, the heterogeneity in the GND spatial distribution in all samples reflects stress and strain heterogeneity at the grain scale and the on-average higher GND densities denote stress and strain concentration in the neck domains of samples stretched by >20% bulk strain.

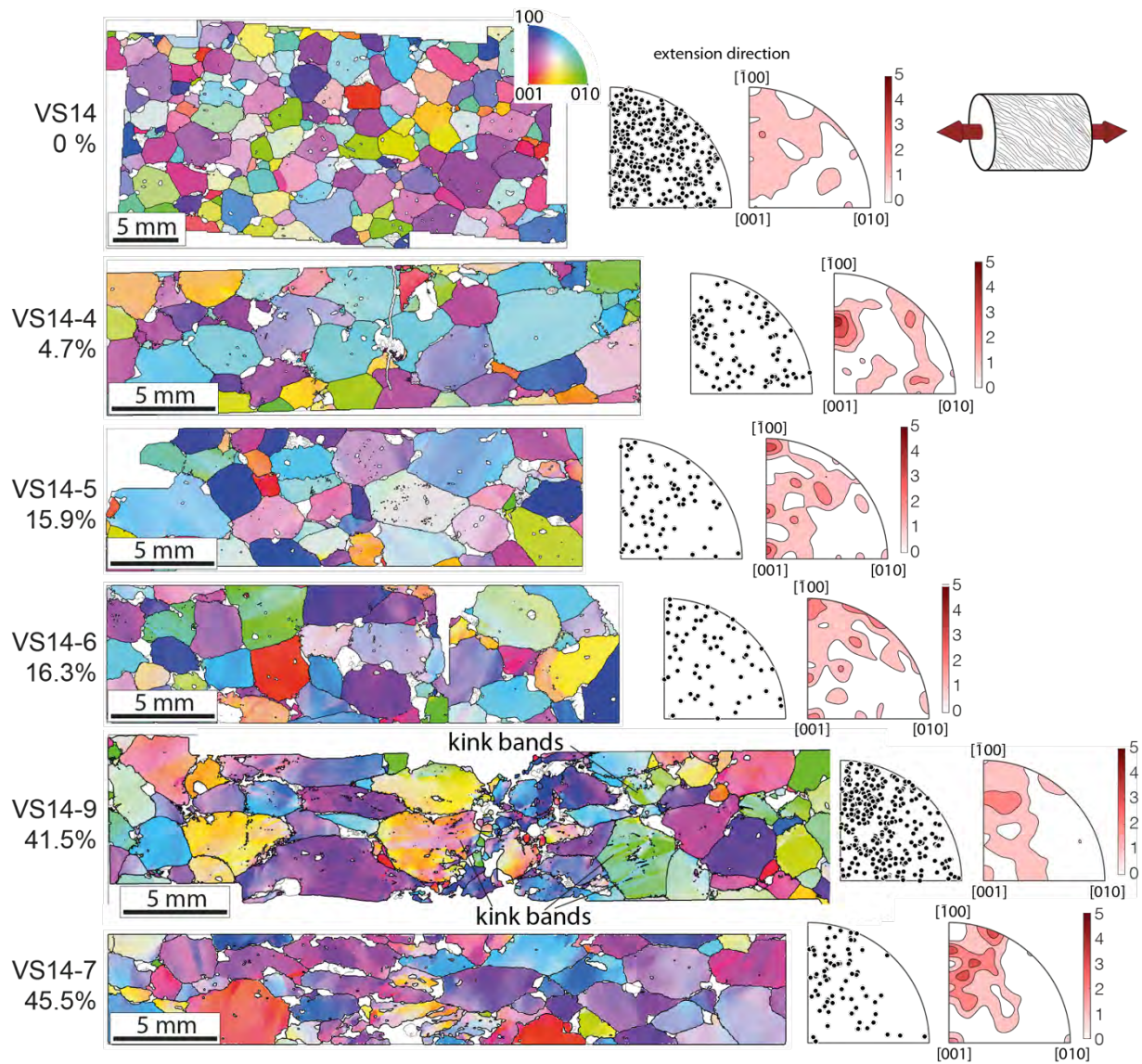

**Fig. 8.** Evolution of the microstructure and of the CPO of olivine with increasing strain illustrated by crystal orientation maps and inverse pole figures (IPF) that indicate the orientation of the bulk extension direction relative to the crystal reference frame for all VS14 samples. Maps are colored as a function of the orientation of the bulk extension direction relative to the crystal reference frame (IPF legend in the insert on the top map). Grain boundaries are displayed in black. Pyroxenes, spinel, dynamically recrystallized olivine grains, and missing material (lost during preparation of the sections for EBSD analyses) are displayed in white. Contours of IPF plots are at one multiple of a uniform distribution intervals. The blotchiness of the contoured IPFs results from sampling a too small number of grains even when analyzing the full specimen cross-section; the maxima mark the orientations of the coarser crystals.

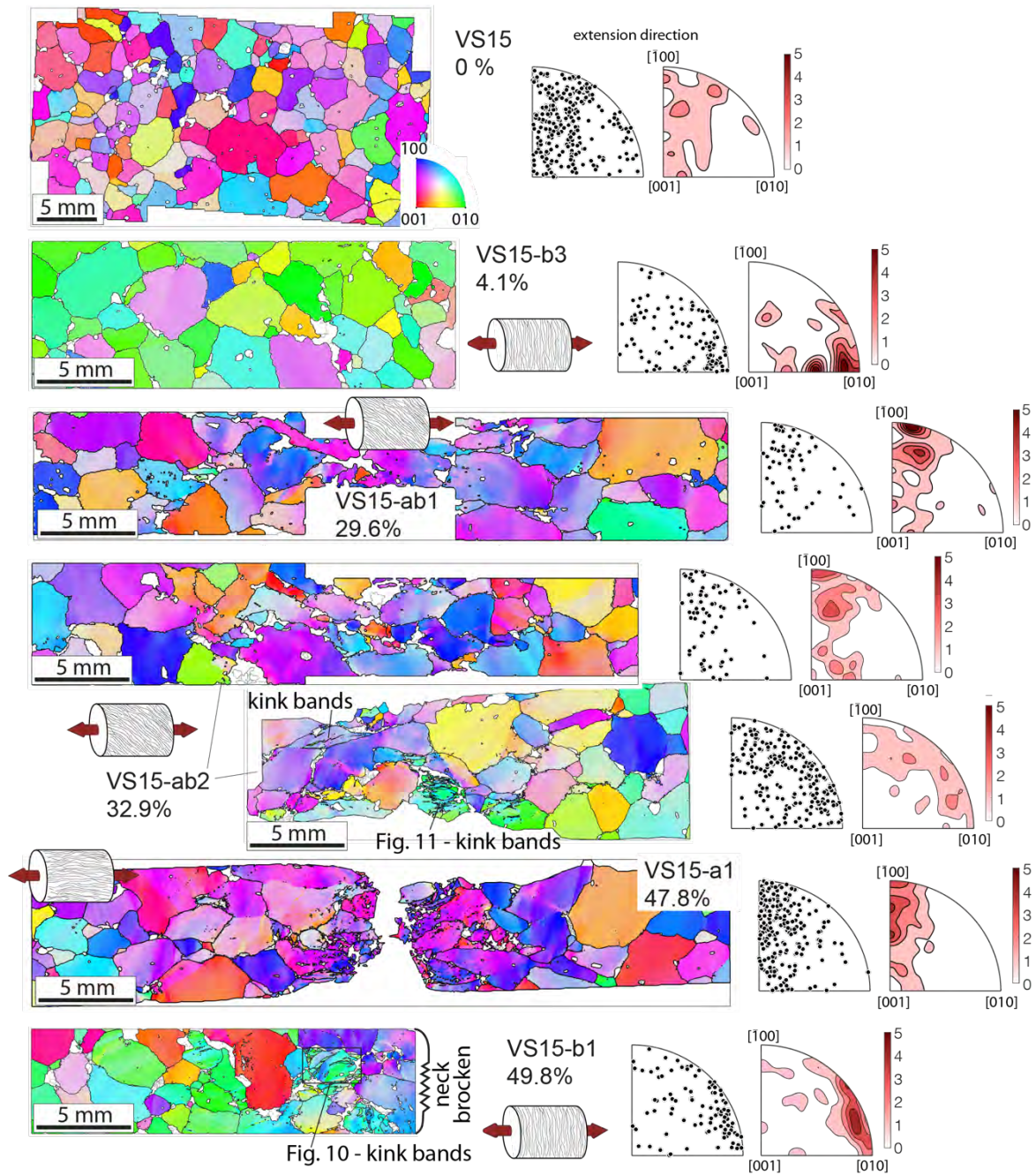

**Fig. 9.** Evolution of the microstructure and of the CPO of olivine with increasing strain illustrated by crystal orientation maps and inverse pole figures (IPF) indicating the orientation of the bulk extension direction relative to the crystal reference frame for VS15 dunites. Maps are colored as a function of the orientation of the bulk extension direction relative to the crystal reference frame (IPF legend as insert on the top map). Grain boundaries are displayed in black. Pyroxenes, spinel, dynamically recrystallized olivine grains, and missing material (lost during preparation of the sections for EBSD analyses) are displayed in white. Contours of IPF plots are at one multiple of a uniform distribution intervals. Small cylinders show the orientation of the initial fabric relative to the imposed extension in the different experiments.

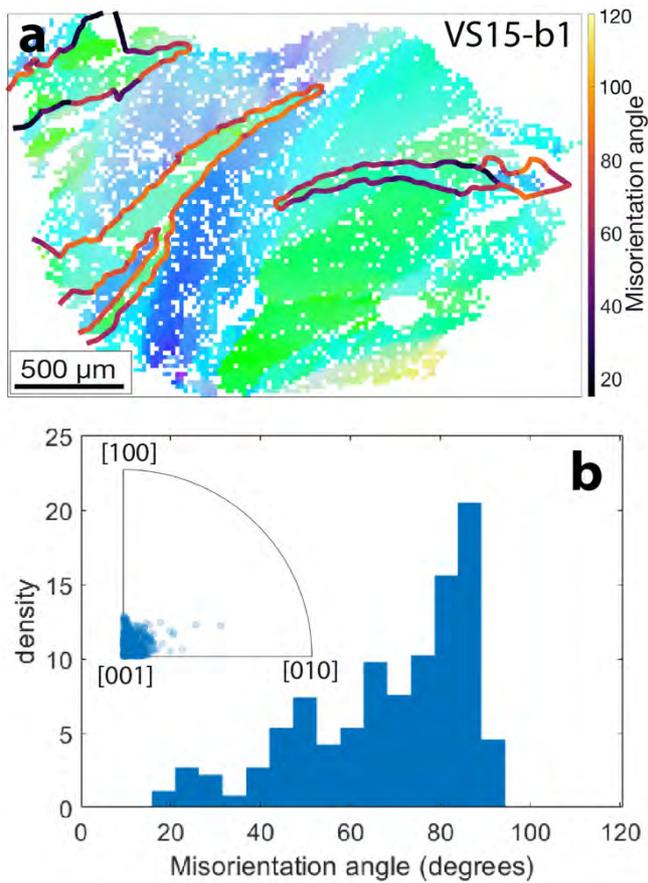

**Fig. 10.** Kinked olivine crystal in coarse-grained dunite VS15-a1 (location indicated in Fig. 9). (a) Orientation map in which kink boundaries with variable misorientation angle are highlighted. (b) Misorientation angle and axes distributions across the kink boundaries highlighted in (a).

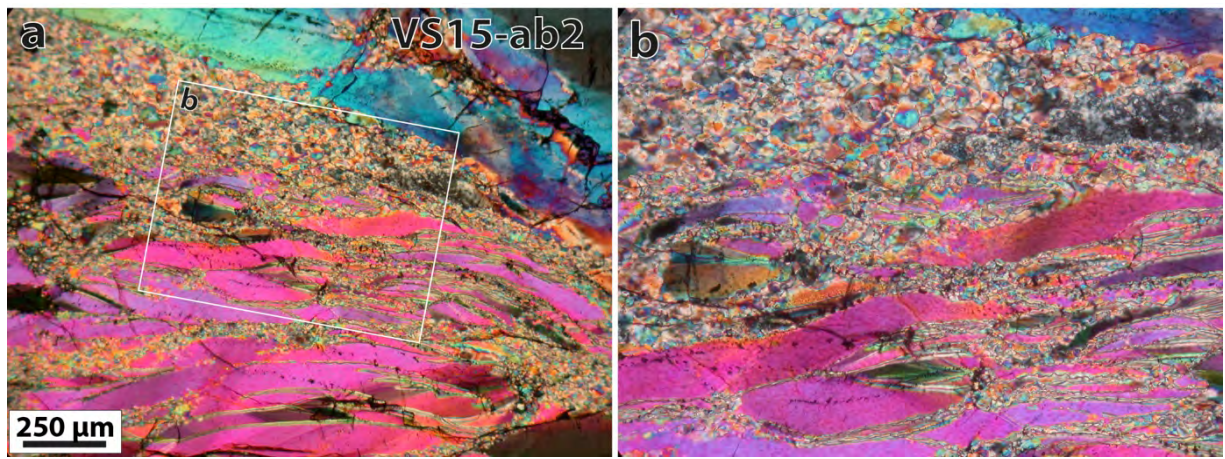

**Fig. 11.** Photomicrographs of a kinked and partially recrystallized olivine crystal in the neck region of coarse-grained dunite VS15-ab2 (location indicated in Fig. 8). Bulk extension direction is parallel to the long side of (a).

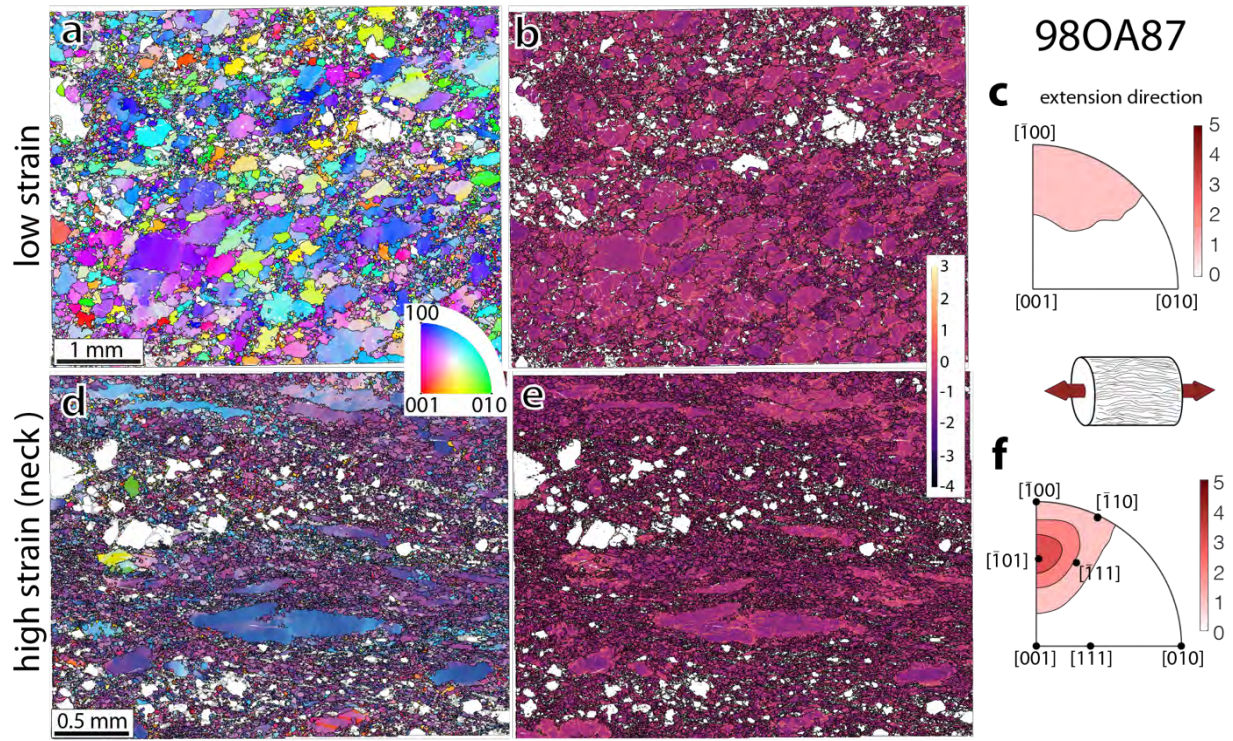

**Fig. 12.** Evolution of the microstructure and the CPO of olivine with increasing strain illustrated by (a and d) crystal orientation, (b and e) KAM maps and (c and f) inverse pole figures (IPF) indicating the orientation of the bulk extension direction relative to the crystal reference frame for (a-c) low strain (close to the piston) and (d-f) high strain (neck) zones of fine-grained harzburgite 90OA87. Orientation maps are colored as a function of the orientation of the bulk extension direction relative to the crystal reference frame (IPF legend in the insert). High KAM values (hotter colors, note the logarithmic scale) delineate the subgrain structure. Grain boundaries are displayed in black. Pyroxenes and spinel are displayed in white. The small cylinder shows the orientation of the initial fabric relative to the imposed extension. In (a and d) the imposed extension is parallel to the long side of the maps.

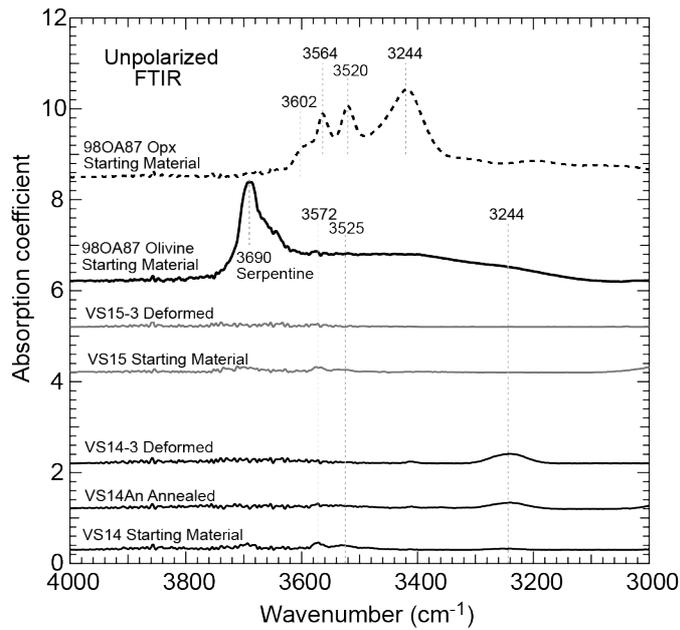

**Fig. 13.** Representative unpolarized infrared spectra for olivine in 98OA87 (starting material), VS14 (starting material), VS14An (annealed at 1200°C and 300 MPa confining pressure), VS14-3 (deformed to 9.1% bulk strain), VS15 (starting material), VS1563 (deformed to 4.1% bulk strain). For the 98OA87 starting material, spectra for olivine were contaminated by the presence of serpentine along grain boundaries. However, clear spectra were obtained for orthopyroxene, allowing the estimation of the hydrogen concentration of olivine based on partition coefficients. Spectra are offset vertically for clarity and are all normalized to 1 cm thickness.

**Table 1. Experimental data**

| Sample | Orientation of the cylinder | Initial length (mm) | Bulk strain (%) | Flow / Peak / Final* stress (MPa, ±5MPa) | Comments |
|---|---|---|---|---|---|
| VS14-3 | In the starting material thin section plane | 21.00 | 9.1 | 123 / 159 / 159 | Force gauge failure |
| VS14-4 | Random | 26.12 | 4.7 | 114 / 114 / 114 | Jacket failure |
| VS14-5 | Random | 23.10 | 14.9 | 123 / 147 / 147 | Furnace failure |
| VS14-6 | Random | 22.18 | 15.4 | 122 / 144 / 144 | Furnace failure |
| VS14-7 | Random | 26.42 | 44.7 | 108 / 156 / 71 | Test successfully completed |
| VS14-8 | In the starting material thin section plane | 25.46 | 39.4 | - | Computer crash |
| VS14-9 | In the starting material thin section plane | 25.05 | 41.4 | 120 / 151 / 151 | Test successfully completed |
| VS15a-1 | // to [100]max | 25.92 | 47.6 | 131 / 183 / 92 | Sample failure |
| VS15b-1 | // to [010]max | 21.65 | 48.1 | 108 / 115 / 20 | Sample failure |
| VS15b-3 | // to [010]max | 22.26 | 4.1 | - | Furnace failure |
| VS15ab-1 | 45° to [100]max and 90° to [010]max | 24.01 | 28.6 | 143 / 156 / 121 | Test successfully completed |
| VS15ab-2 | 45° to [100]max and 45° to [010]max | 24.63 | 33. | 118 / 170 / 150 | Test successfully completed |
| 90OA87 | At low angle to the preexisting lineation | 24.75 | 49.7 | 165 / 165 / 73 | Test successfully completed |

*Final stress is an apparent value, not corrected for the effects of heterogenous deformation

**Table 2. Dynamically recrystallized grain size of olivine and differential stresses estimated using the Van der Wal et al. (1993) piezometer**

| Sample | n | arith. mean | Conf. Interval at 95%[a] | | | | Stress (MPa) | Conf. Interval at 95% | | | |
|---|---|---|---|---|---|---|---|---|---|---|---|
| | | | lower | upper | abs err ($\pm$) | c.v. ($\pm$ %)[b] | | lower | upper | lower (%) | upper (%) |
| vs14-7la | 615 | 12.2 | 11.7 | 12.8 | 0.5 | 4.3 | 167.4 | 162.2 | 173.1 | 3.1 | 3.4 |
| vs14-7rb | 1753 | 12.8 | 12.5 | 13.1 | 0.3 | 2.3 | 162.2 | 159.5 | 165.0 | 1.7 | 1.7 |
| vs14-9 | 1153 | 11.7 | 11.4 | 12.0 | 0.3 | 2.6 | 172.7 | 169.4 | 176.2 | 1.9 | 2.0 |
| vs15ab1 | 1625 | 12.7 | 12.4 | 13.0 | 0.3 | 2.3 | 162.8 | 160.2 | 165.7 | 1.6 | 1.8 |
| vs15ab2 | 1259 | 10.7 | 10.4 | 10.9 | 0.3 | 2.6 | 185.5 | 182.0 | 189.2 | 1.9 | 2.0 |
| vs15ab2-bis | 2217 | 11.6 | 11.4 | 11.9 | 0.2 | 2.1 | 174.0 | 171.2 | 176.7 | 1.6 | 1.6 |
| vs15ab2-ter | 330 | 11.4 | 10.8 | 12.0 | 0.6 | 5.1 | 176.5 | 170.0 | 183.5 | 3.7 | 4.0 |
| vs15b1 | 849 | 12.7 | 12.2 | 13.1 | 0.5 | 3.5 | 163.2 | 159.0 | 167.7 | 2.6 | 2.8 |
| 98OA87-neck | 24711 | 8.6 | 8.5 | 8.6 | 0.1 | 0.7 | 219.0 | 217.6 | 220.1 | 0.6 | 0.5 |
| 98OA87-low strain | 12742 | 15.9 | 15.7 | 16.1 | 0.2 | 1.2 | 137.8 | 136.5 | 138.9 | 0.9 | 0.9 |

[a]*Confidence intervals for the mean at 95% of certainty using the standard error formula with the t-score (ASTM International, 2013; Lopez-Sanchez, 2020)*
[b]*Coefficient of variation: Error relative to the arithmetic mean in percentage (100 x abs err / mean)*

**Table 3. FTIR data**

| Sample | Thickness (μm) | Integrated Absorbance ($cm^2$) | H concentration Paterson (1982) calibration ppm $H_2O$ wt. | H concentration Withers et al. (2012) or *Bell et al. (1985) calibration ppm $H_2O$ wt. |
|---|---|---|---|---|
| 98OA87 (starting material) olivine | 445 | *Contaminated by serpentine* | | |
| 98OA87 (starting material) opx | 416 | 432 | 76 | 87* |
| VS14 (starting material) olivine | 589 | 11.8 | 2 | 4 |
| VS14An (annealed) olivine | 560 | 15.1 | 2 | 5 |
| VS14-3 (deformed) olivine | 421 | 16.6 | 2 | 6 |
| VS15 (starting material) olivine | 455 | 2 | <0.5 | 1 |
| VS15-b3 (deformed) olivine | 507 | 0.5 | 0 | 0.5 |

# Supplementary Material to

# Deformation of upper mantle rocks with contrasting initial fabrics in axial extension

Walid Ben Ismail[1#], Andréa Tommasi[2*], Marco Antonio Lopez[2], Ernest H. Rutter[1], Fabrice Barou[2], Sylvie Demouchy[2]

[1]*Rock Deformation Laboratory, Dept. Earth Sciences, University of Manchester, UK*
[2]*Géosciences Montpellier, CNRS & Université de Montpellier, Montpellier, France*

This supplementary material is composed of:
- The detailed presentation of the FTIR analysis method
- Fig. S1 illustrating isolated partial melting pockets due to the local presence of hydrous phases along grain boundaries in the deformed samples.
- Fig. S2 showing details of the kink bands.
- Table S1 presenting the intragranular misorientation data extracted from the EBSD maps and presented in Fig. 5 of the main text.
- The presentation of the viscoplastic self-consistent simulations performed to isolate the effect of the olivine CPO on the mechanical behavior of the samples, which includes Table S2 and Figure S3.

**Fourier-transform infrared spectroscopy analyses**

Slabs of the three starting materials (90OA87, VS14, VS15), one annealed sample (VS14An) and two selected post-deformation samples (VS14-3, VS15-b3) were hand-polished using a polishing jig and diamond-lapping films with grit sizes ranging from 30 to 1 μm. The final thickness of each slab is reported in Table 3. The samples impregnated by epoxy during preparation for EBSD analyses were not suitable for FTIR as the glue contains OH group.

Hydroxyl groups in olivine were analyzed using transmission Fourier-transform infrared (FTIR) spectroscopy at the Laboratoire Charles Coulomb (University of Montpellier, France). Unpolarized mid-infrared spectra were acquired using a Bruker INVENIO R (KBr/Ge beam splitter) coupled to a Bruker HYPERION microscope equipped with a liquid nitrogen-cooled MCT detector (Mercatel alloy, HgCdTe). Unpolarized IR measurements were performed following the protocol of Denis et al. (2015). Optically clean areas were chosen for analyses, and 200 scans were accumulated at a resolution of 4 cm$^{-1}$ using a square aperture of different sizes (50 × 50 μm, 100 × 100 μm, or 200 × 200 μm) as a function of the size of the clean area available. Each spectrum was baseline corrected and the absorbance normalized to 1 cm thickness. The samples were cleaned using acetone bath several times prior to analyses and placed on a CaF$_2$ plate when necessary. A minute amount of Fluorolube (ThermoScientific$^{TM}$) was used to eliminate thickness fringes when needed.

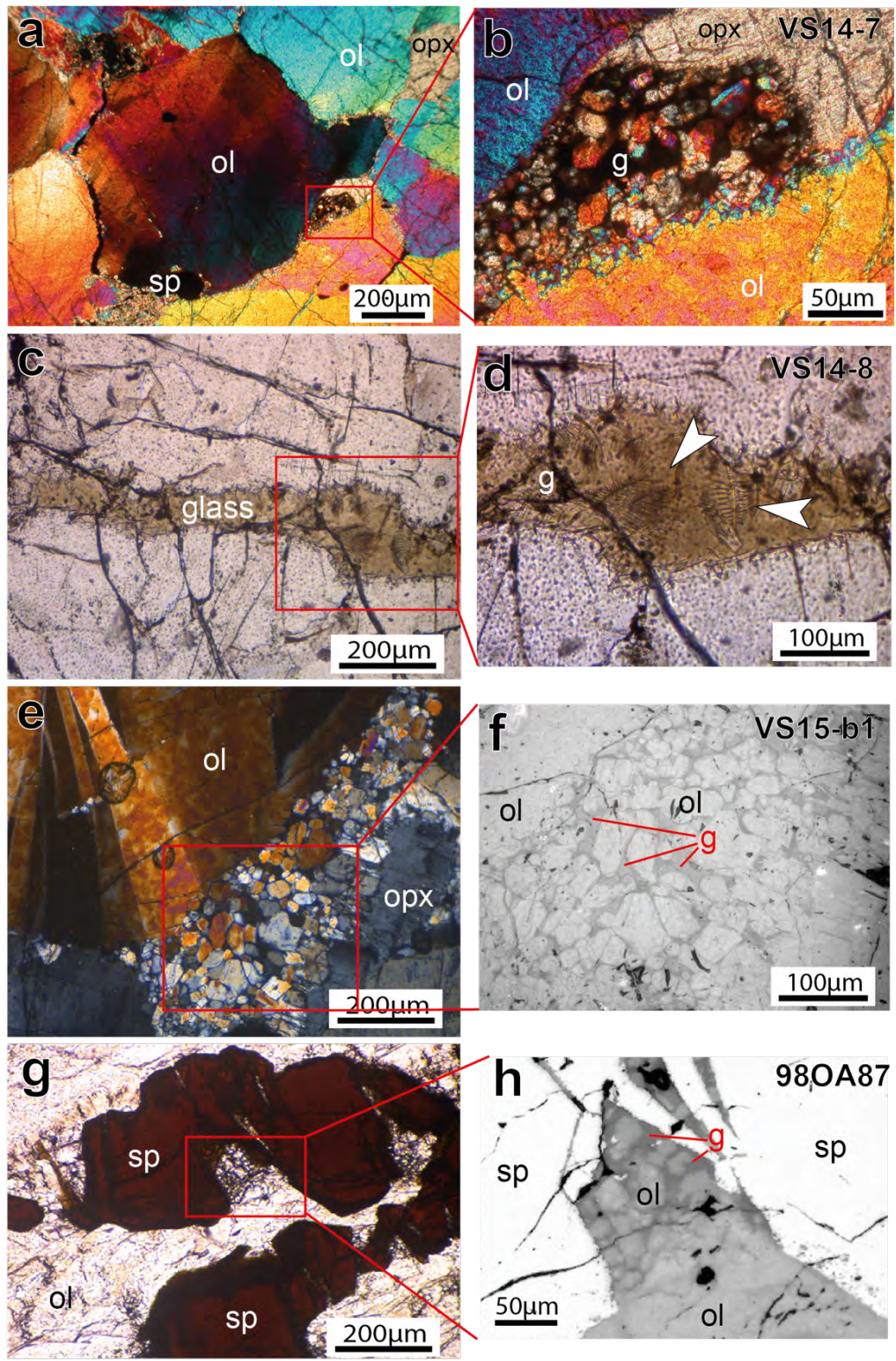

**Fig. S1.** Evidence for partial melting in different deformed samples. (a,b) «Melt» pocket at the contact with an orthopyroxene composed of small olivine crystals enclosed in glassy material (g) in sample VS14-7. (b,c) The largest glassy material pocket observed localized in a olivine-olivine grain boundary normal to the bulk extension direction in sample VS14-8. Arrows indicate feather like structures in the glass. (e,f) «Melt» pocket at the contact with an orthopyroxene composed of small olivine crystals enclosed in glassy material in sample VS15-b1. (g,h) «Melt» pocket at the contact with a spinel grain composed of small olivine crystals enclosed in glassy material in sample 98OA87. All images are photomicrographs: (a,b,e) cross-polars, (c,d,f) natural light, (f,g) reflected light.

Fig.S2: Crystal orientation maps colored as a function of the orientation of the crystals relative to the extension direction (IPF legend) and misorientation profiles (relative to the initial point) across kink bands.

**Effect of the olivine CPO on the mechanical behavior of the samples assessed by viscoplastic self-consistent simulations**

We compare the instantaneous mechanical response of olivine polycrystals with the olivine CPO measured in the low strain and shadow zone subjected to axial extension in the same orientation as in the experiments to the strength predicted for an olivine polycrystal with an initially isotropic CPO subjected to up to 100% axial extension. We also compare the instantaneous mechanical response of olivine polycrystals with the olivine CPO measured in samples VS15a1, VS15b1, VS15ab1, VS15ab2 to check if the observed differences in initial strength may be explained by the mechanical anisotropy due to the orientation of the olivine CPO of each sample relative to the imposed extension.

The mechanical response of the olivine polycrystals was modeled using the second-order formulation of the VPSC model. This approach considers heterogeneous stress and strain rate within the grains and accounts for intragranular strain rate fluctuations to predict the effective polycrystal behavior (Ponte-Castañeda et al., 2002; Lebensohn et al., 2011). Individual crystals deform by dislocation glide on a finite number of slip systems, which are a function of the crystal structure. The relative strength of these slip systems, that is, the critical resolved shear stress (CRSS) needed to activate dislocation glide on each system, depends on the temperature, pressure, and stress. In the present study, we used CRSSs (Table S3) derived from deformation of natural olivine single crystals ($Fo_{90}$) at high temperatures and low pressure (Bai et al., 1991). As the aim of the present simulations is solely to evaluate the effect of the CPO on the mechanical behavior, hardening due to microstructural evolution (dislocation entanglements) is not considered in the simulations.

**Table S2:** Slip systems data used in the VPSC simulations

| Slip Systems | Critical Resolved Shear Stress# | Stress exponent |
|---|---|---|
| (010)[100] | 1 | 3 |
| (001)[100] | 1.5 | 3 |
| (010)[001] | 2 | 3 |
| (100)[001] | 3 | 3 |
| (011)[100] | 4 | 3 |
| (110)[001] | 6 | 3 |
| {111}<110>[a] | $\beta$ | 3 |
| {111}<011>[a] | $\beta$ | 3 |
| {111}<101>[a] | $\beta$ | 3 |

\* Lower bound simulations only consider the first four slip systems
\# Adimensional values; normalized by the flow stress of the (010)[100] slip system
[a] dummy slip systems, which represent additional strain accommodation processes and do not contribute to plastic spin

Mixed boundary conditions were enforced because they better mimic the actual conditions in the laboratory experiments: imposed axial extension along and homogeneous stress in the plane normal to the piston axis.

$$L = \begin{bmatrix} * & 0 & 0 \\ 0 & * & 0 \\ 0 & 0 & 1 \end{bmatrix}; \sigma = \begin{bmatrix} -1 & * & * \\ 0 & -1 & * \\ * & * & * \end{bmatrix}$$

\* = not imposed

The relative strength of the VS15 samples subjected to axial extension in different directions relative to the olivine CPO (Table S3) is consistent with the measured strengths, except for sample VS15-b1, which should have presented a higher yield strength, on the same order than samples VS15-a1 and VS15-ab1. Indeed, analysis of the olivine orientation maps for VS15-b1 (Fig. 9 in the main text) shows that deformation in this sample was concentrated in a zone with an anomalous orientation, where most crystals had their [100] at low angle to the imposed extension.

The olivine CPO in the neck zone of sample 98OA87, where local strains of up to 100% were estimated (Fig. 3 in the main text), differs significantly to that predicted for an olivine polycrystal with an initially random CPO accommodating up to 100% axial extension by dislocation glide: it is less concentrated and the maximum concentration of [100] is at 16° to the extension direction (Fig. S3). This implies that the CPO evolution in the experiments is significantly modified by dynamic recrystallization. This change in the CPO evolution avoids the hardening that should occur due to the progressive rotation of the [100] towards the extension direction, as the neck region remains weaker than an isotropic aggregate, instead of 1.5 times harder (Fig. S3).

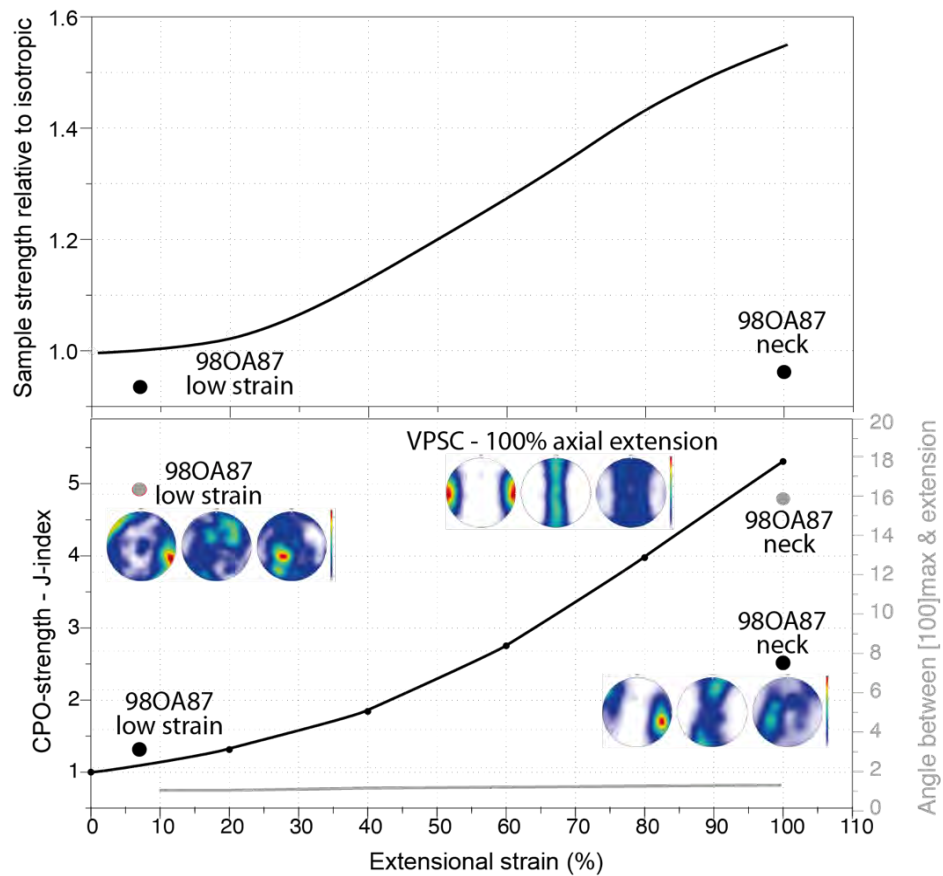

**Figure S3.** Evolution of the sample strength (top) and olivine CPO (bottom) as a function of the extensional strain predicted by the VPSC approach for an aggregate with an initially isotropic CPO subjected to axial extension (full

lines) and measured in a low strain and the neck regions of the sample 98OA87 (full symbols). The olivine CPO evolution is characterized by the change in CPO strength (quantified by the J-index, which is the integral of the squared orientation distribution function, in black) and by the angle between the [100] maximum and the extension direction (in gray). The inserts show pole figures of the CPO (extension direction is E-W) in a low strain and the neck regions of the sample 98OA87 and at the end of the simulation.

**Table S3**: Summary of the results of all VPSC simulations

| Initial olivine CPO | Strain (%) | Stress* | J-index | Angle** | Strength contrast |
|---|---|---|---|---|---|
| 98OA87_neck | 100 | 1.62E+01 | 2.5156 | 15.9481 | 9.62E-01 |
| 98OA87_shadow | 7 | 1.57E+01 | 1.3153 | 16.2901 | 9.35E-01 |
| VS15a1 | 0 | 1.92E+01 | 3.1901 | 14.0875 | 1.14E+00 |
| VS15ab1 (45a/90b) | 0 | 1.95E+01 | 5.1896 | 8.0316 | 1.16E+00 |
| VS15ab2 (45a/45b) | 0 | 1.58E+01 | 1.3719 | 29.7466 | 9.36E-01 |
| VS15b1 | 0 | 1.87E+01 | 4.1774 | 57.664 | 1.11E+00 |
| random | 0 | 1.68E+01 | 1 | n/d | 1.00E+00 |
| random | 20 | 1.73E+01 | 1.3132 | 0.8478 | 1.03E+00 |
| random | 40 | 1.90E+01 | 1.8421 | 0.9364 | 1.13E+00 |
| random | 60 | 2.15E+01 | 2.7522 | 0.9661 | 1.28E+00 |
| random | 80 | 2.41E+01 | 3.9754 | 0.9776 | 1.43E+00 |
| random | 100 | 2.60E+01 | 5.3078 | 0.982 | 1.55E+00 |

\* Normalized to the CRSS of the [100](010) slip system
\*\* Angle between the [100]max and the extension direction
\*\*\* Relative to the isotropic behavior